\documentclass{biophys-new}
\usepackage[utf8]{inputenc}
\usepackage{graphicx}
\usepackage[colorlinks,allcolors=cyan!70!black]{hyperref}
\usepackage{dcolumn}   
\usepackage{bm}        
\usepackage{amssymb}   
\usepackage{amsmath}
\usepackage[Euler]{upgreek}
\usepackage{float}
\usepackage{multirow}
\usepackage{booktabs}
\usepackage{textcomp}
\usepackage[Euler]{upgreek}
\usepackage{lipsum}
\usepackage{rotating}
\usepackage{tikz}
\usepackage[normalem]{ulem}

\DeclareMathAlphabet\mathbfcal{OMS}{cmsy}{b}{n}
\newcommand{\bcal}[1]{\bf{\mathbfcal{#1}}}
\title{Computing 3D chromatin configurations from contact probability maps by Inverse Brownian Dynamics}
\runningtitle{Biophysical Journal Template} 

\author{K. Kumari, B. D{\"u}nweg, R. Padinhateeri, J. R. Prakash}
\runningauthor{Author1 and Author2} 

\papertype{Article}

\begin{document}

\begin{frontmatter}

\begin{abstract}
The three-dimensional organization of chromatin, on the length scale of a few genes, is crucial in determining the functional state --- accessibility and the amount of gene expression --- of the chromatin. Recent advances in chromosome conformation capture experiments provide partial information on the chromatin organization in a cell population, namely the contact count between any segment pairs, but not on the interaction strength that leads to these contact counts. However, given the contact matrix, determining the complete 3D organization of the whole chromatin polymer is an inverse problem.       
In the present work, a novel Inverse Brownian Dynamics (IBD) method based on a coarse grained bead-spring chain model has been proposed to compute the optimal interaction strengths between different segments of chromatin such that the experimentally measured contact count probability constraints are satisfied. 
Applying this method to the $\alpha$-globin gene locus in two different cell types, we predict the 3D organizations corresponding to active and repressed states of chromatin at the locus. 
We show that the average distance between any two segments of the region has a broad distribution and cannot be computed as a simple inverse relation based on the contact probability alone. 
The results presented for multiple normalization methods suggest that all measurable quantities may crucially depend on the nature of normalization. We argue that by experimentally measuring predicted quantities, one may infer the appropriate form of normalization.  
\end{abstract}

\begin{sigstatement}
Chromosome conformation capture experiments such as 5C and Hi-C provide information on the contact counts between different segments of chromatin, but not the interaction strengths that lead to these counts. Here a methodology is proposed by which this inverse problem can be solved, namely, given the contact probabilities between all segment pairs, what is the pair-wise interaction strength that leads to this value? With the knowledge of pair-wise interactions determined in this manner, it is then possible to evaluate the 3D organization of chromatin and to determine the true relationship between contact probabilities and spatial distances. 
\end{sigstatement}
\end{frontmatter}

\section{Introduction}

Even though all the cells in multi-cellular organisms have the same DNA sequence, they function differently based on the cell type. For example, the phenotype of a skin cell is significantly different from that of a neuronal cell~\citep{Alberts2015,ecker2012genomics}.
One of the important factors for this variation is  hypothezised to be the three-dimensional organization of DNA inside the cell nucleus and its variability from cell-type to cell-type~\citep{larson2018role,gilbert2005chromatin,fraser2007nuclear,bickmore2013spatial}. While findings of the recent chromosome configuration capture experiments  (3C, 4C, 5C, Hi-C)~\citep{Dekker1306,Simonis2006,Dostie2006,Lieberman-Aiden289} lend credence to this hypothesis, the outcomes of these experiments are frozen snapshots of a sparse set of points along DNA that do not give a complete understanding of the  3-dimensional organization of the genome. In this paper, a methodology based on a coarse-grained polymer model for DNA is proposed, which enables the unravelling of its spatio-temporal organization that is consistent with experimentally observed contact maps.

The complex folding of meter-long DNA into micrometer-sized chromosome, with topologically associated domains (TAD) and contact domains, has been revealed at few kb resolution by state of the art Hi-C experiments~\citep{Nora2012,Dixon2012,nora2013,Rao20141665,rowley2018organizational,mir2019chromatin}.
More insight into the role played by the 3D organization of the genome in the functioning of a cell on the length-scale of genes, is provided by 3C and 5C experiments~\citep{sanyal2012long}. Essentially, all these chromatin conformation capture experiments lead to information on the count of contacts between any pair of segments along the DNA chain backbone, represented in the form of contact (``heat'') maps.

Several  attempts have been made to understand the 3D organization of the genome using a variety of techniques developed previously to understand the statics and dynamics of polymers~\citep{Schiessel_2003,teif2011condensed,ganai2014chromosome,tamar2019pnas,dans2016multiscale,Gehlen2012,yan2005statistics,DiPierro2016,Mirny2011,Bancaud2012,rosa2008structure,di2017novo,jost2014modeling,gursoy2017computational}. 

Early models focused on understanding the non-equilibrium nature of chromatin organization and the polymer physics behind large scale packaging~\citep{Mirny2011,Bancaud2012,rosa2008structure}.
Subsequent studies that focus on reconstructing the 3D structure from the contact maps are predominantly based on assuming that there is a direct correlation between the magnitude of the contact count and the spatial distance between the relevant pairs \citep{Fraser2009,Duan2010,Tanizawa2010,rousseau2011three,Bau2011a,Paulsen2017,Paulsen2018}. These investigations have led to important insights about the 3D consequences of differences present in the contact maps such as the spatial organization of ON and OFF states of certain genes. However, all these efforts have certain limitations.
As mentioned above, nearly all the computational studies convert contact counts obtained from Hi-C experiments into spatial distances, using a pre-decided formula. That means, given a contact count matrix, such methods do not predict the distances between different chromatin segments, rather they take the distance values as inputs, based on certain assumptions.
They then use conventional Monte Carlo (or equivalent) methods to find steady state configurations of the chromatin, given a distance map between different DNA segment pairs. In other words, the existing models consider this as a ``forward'' problem of computing equilibrium configurations of chromatin as a consequence of assuming certain spatial distance between bead-pairs.  However, the problem of computing 3D configurations of a  chromatin polymer, given a contact map, is not a ``forward'' problem but rather an ``inverse'' problem~\citep{Meluzzi2013}. The question is, given a contact map, what are the optimal interactions between different segments of chromatin such that the experimentally seen contact map emerges. To the best of our knowledge, no study exists that solves chromatin configurations of genes, considering it as an inverse problem. 
Another shortcoming is that the experimentally obtained contact counts are not converted to ``absolute'' contact probabilities. Some of the existing methods remove various systematic biases and convert the contact counts to relative contact probabilities. Some of these techniques are: iterative correction and eigenvector decomposition (ICE)~\cite{imakaev2012iterative}, sequential component normalization (SCN)~\cite{cournac2012normalization}, Knight-Ruiz (KR)~\cite{knight2013fast}, chromoR~\cite{shavit2014combining}, multiHiCcompare~\cite{stansfield2019multihiccompare}, and HiCNorm~\cite{hu2012hicnorm}.
In the current work, we examine the existing ICE normalization method and compare it with a method processed here based on a simple process of converting contact counts to contact probabilities though a parameter representing the ensemble size. 
We show that the structural properties of the gene loci depend on the precise values of contact probabilities.
It should also be noted that all prior efforts are based on Monte Carlo methods, and hence they cannot predict the dynamics of chromatin --- they only obtain information on static configurations of the genome. 

In summary, while current models have made important progress in constructing 3D structure from the contact maps, they suffer from one or more of the following shortcomings:
\begin{enumerate}
	
	\item An \textit{a priori} assumption regarding the probability of contact between pairs of segments and their spatial distances.
	
	\item The introduction of harmonic springs between interacting pairs that implies an attractive force between these pairs that does not decay with distance, but rather increases.
	
	\item The use of simulation methods that are limited to providing information on static configurations.
	
	\item Considering the problem of computing 3D configurations as a ``forward'' problem, with no attempt to determine the interaction strengths between segment pairs that lead to 3D structures that are consistent with observed contact maps.
	
	\item The failure to obtain an accurate representation of dynamic behaviour by failing to include hydrodynamic interactions~\cite{prakash2019universal} between segment pairs.
	
\end{enumerate}

In this work, a methodology is introduced that addresses all these shortcomings.  Chromatin on the length-scale of a gene is represented by a coarse-grained bead spring chain polymer model, with a  potential of interaction between pairs of beads that can be tuned to accommodate varying strengths of interaction. A Brownian dynamics simulation algorithm that includes hydrodynamic interactions, and an iterative scheme based on Inverse Monte Carlo is developed that enables the generation of 3D configurations that are consistent with the contact maps.  This methodology is then applied to obtain the static 3D configurations from 5C contact maps of the $\alpha$-globin gene locus, both in the ON and OFF states of the gene. Further, since hydrodynamic interactions are taken into account, the approach has the potential to examine the dynamic transitions between the ON and OFF states. In the present work, however, since the focus is on reproducing heatmaps and generating 3D configurations (which are both static properties), dynamic properties have not been considered.

The outline of the paper is as follows. The key governing equations of the model and the simulation algorithm are summarised in section~\ref{sec:model}. In section~\ref{sec:ibd}, the inverse Brownian dynamics method is introduced in a general context. The validation of the proposed approach with the help of a prototype is presented in section~\ref{sec:val}. 
The coarse-graining procedure used here is described in section~\ref{sec:cg}.
Resolution of the issue of determining the contact probabilities from contact counts is proposed in section~\ref{sec:c2p}. Results for the static 3D configurations of $\alpha$-globin locus are discussed in section \ref{sec:3d_confi}, while the relationship between spatial distances and contact probabilities is highlighted in section~\ref{sec:c2d}. The principal conclusions of this work are summarized in section~\ref{sec:conclusion}.

\section{Model and Methods}
Chromosome conformation capture experiments such as 5C and Hi-C provide information about the contact counts between different segments of chromatin, but not the interaction strengths that lead to these counts. Here, we propose a methodology by which this inverse problem can be solved, namely, given the contact probabilities between all segment pairs, what is the pair-wise interaction strength that leads to this value? Additionally, the fact that experiments only give contact counts and not probabilities need to be dealt with. In section~\ref{sec:model}, we first provide the principal governing equations and the details of the interactions. 
In section~\ref{sec:ibd}, we describe the inverse Brownian dynamics algorithm by which the interaction strengths $\epsilon_{\mu\nu}$ can be estimated given the set of contact probabilities $p_{\mu\nu}$.
\subsection{\label{sec:model} Polymer model}
To compute the 3-dimensional organization of the genome, the chromatin is coarse-grained into a bead-spring chain of $N$ beads connected by $N-1$ springs. The chain configuration is specified by the set of position vectors of the beads $\bm r_{\mu}(\mu = 1, 2, ..., N)$. For simulation purposes, all distances are made dimensionless by using the characteristic length scale $l_0 = \sqrt{k_{\rm B}T/k_s}$ arising from the ratio of thermal energy---where $k_{\rm B}$ is the Boltzmann constant, $T$ is the temperature---and the spring constant $k_s$. Throughout this manuscript the asterisk superscript is used to indicate dimensionless quantities ($ r_{\mu}^* =  r_{\mu}/l_0$). The adjacent beads in the polymer chain are bonded via a Fraenkel spring, with a non-dimensional spring potential $U_{\mu}^{\rm s*}$ between bead $\mu$ and $(\mu+1)$, given by
\begin{align}\label{eq:spring}
U_{\mu}^{\rm s*}= \frac{1}{2}\left[ \left( r_{\mu+1}^* -r_{\mu}^*\right)- r_0^*\right]^2
\end{align}
where $ (r_{\mu+1}^* -r_{\mu}^*)$ is the non-dimensional distance between bead $\mu $ and $\mu+1$, and $r_0^*$ is the dimensionless natural length of the Fraenkel spring. 
To mimic protein-mediated interactions between different parts of the chromatin polymer, it is necessary to introduce a potential energy function. Typically, this is achieved with a Lennard-Jones (LJ) potential or with harmonic spring interactions~\citep{Meluzzi2013}. However, in the present study, the following non-dimensional Soddemann-Duenweg-Kremer (SDK)~\citep{Soddemann2001} potential is introduced between any two non-adjacent beads $\mu$ and $\nu$,
\begin{align}\label{eq:sdk}
U_{\mu\nu}^{\textrm{SDK*}}=\left\{
\begin{array}{l l l}
&4 \left[ \left( \dfrac{\sigma^*}{  r_{\mu\nu}^*} \right)^{12} - \left( \dfrac{\sigma^*}{  r_{\mu\nu}^*} \right)^6 + \frac{1}{4} \right] - \epsilon_{\mu\nu};\hphantom{{}+{}}  & r_{\mu\nu}^*\leq 2^{1/6}\sigma^* \\ [7pt]
& \dfrac{1}{2} \epsilon_{\mu\nu} \left[ \cos \left(\alpha \dfrac{ r_{\mu\nu}^*}{\sigma^*}+ \beta\right) - 1 \right] ;& 2^{1/6}\sigma^* \leq  r_{\mu\nu}^* \leq r_{\rm c}^*  \\ [7pt]
& 0. &   r_{\mu\nu}^* \geq  r_{\rm c}^*
\end{array}\right.
\end{align}
Here $ r_{\mu\nu}^*= (r_{\mu}^* -r_{\nu}^*)$ is the non-dimensional distance between beads $\mu$ and $\nu$, $\epsilon_{\mu\nu}$ is an independent parameter to control the bead-bead attractive interaction strength between beads $\mu$ and $\nu$ and $2^{1/6}\sigma^*$ represents the minima of the potential where $U_{\mu\nu}^{\textrm{SDK*}} = \epsilon_{\mu\nu}$. The SDK potential has the following advantages compared to the LJ potential: (i) The repulsive part of the SDK potential ($ r_{\mu\nu}^*\leq 2^{1/6}\sigma^*$) representing steric hindrance remains unaffected by the choice of the parameter $\epsilon_{\mu\nu}$. (ii) Protein-mediated interactions in chromatin are like effective ``bonds" formed and broken with a finite range of interaction. Unlike the LJ potential, the SDK potential has a finite attractive range ---the SDK potential energy smoothly reaches zero at the cut off radius, $r_{\rm c}^*$, whose value is set by the choice of  two parameters $\alpha$ and $\beta$. The parameters $\alpha$ and $\beta$ are determined by applying the two boundary conditions, namely, $U_{\mu\nu}^{\text{SDK}} = 0$ at $r_{\mu\nu}^*=r_{\rm c}^*$ and $U_{\mu\nu}^{\text{SDK}}=-\epsilon_{\mu\nu}$ at $r_{\mu\nu}^*=2^{1/6}\sigma^*$. The appropriate choice of the cut-off radius $r_{\rm c}^*$ has been investigated extensively in a recent study~\citep{C9SM01361J}
and it has been shown that a value of $r_{\rm c}^*=1.82\sigma^*$ leads to an accurate prediction of the static properties of a polymer chain in poor, theta and good solvents. The same value is adopted here in the present study.

Given a set of values $\epsilon_{\mu\nu}$ and an initial configuration of the bead-spring chain, the time evolution of the configurations of the polymer chain is evaluated using Brownian Dynamics simulations~\citep{Ottinger1996}, which is a numerical method for solving the following Euler finite difference representation of the stochastic differential equation for the bead position vectors, 
\begin{align}\label{eq:sde_eq}
\bm r_\mu^*(t^* + \Delta t^*) =\bm r_\mu^*(t^*) + \frac{\Delta t^*}{4}  \sum\limits_{\nu=1}^N \bm D_{\mu\nu}\cdot (\bm F_\nu^{s*}+ \bm F_\nu^{\textrm{SDK}*}) +\frac{1}{\sqrt{2}}\sum_{\nu=1}^N \bm B_{\mu\nu} \cdot \Delta \bm  W_\nu
\end{align}
Here $t^* = t/\lambda_0$ is the dimensionless time, with $\lambda_0$ = $\zeta/4k_s$ being the characteristic time scale, in which $\zeta = 6\pi\eta a$ is the Stokes friction coefficient of a spherical bead, $\eta$ is the solvent viscosity and $a$ is the bead radius. $\bm F_\nu^{s*}$ and $\bm F_\nu^{\textrm{SDK}*}$ are the non-dimensional spring and interaction forces computed from the respective potential energy functions provided in Eqs.~(\ref{eq:spring}) and (\ref{eq:sdk}). $\Delta \boldsymbol W_\nu$ is a non-dimensional Wiener process with mean zero and variance $\Delta t^*$ and $\bm{\bm{B}}_{\mu\nu }$ is a non-dimensional tensor whose presence leads to multiplicative noise~\citep{Ottinger1996}. Its evaluation requires the decomposition of the diffusion tensor  
$\boldsymbol D_{\mu\nu}$ defined as $\boldsymbol D_{\mu\nu} = \delta_{\mu\nu} \boldsymbol \delta + \boldsymbol \Omega_{\mu\nu}$, where $\delta_{\mu\nu}$ is the Kronecker delta, $\bm \delta$ is the unit tensor, and $\bm \Omega_{\mu\nu} = {\bm{\Omega}} ( \bm r_\mu^* - \bm r_\nu^*  )$ is the hydrodynamic interaction tensor. 
Defining the matrices $\bcal{D}$ and $\bcal{B}$ as block matrices consisting of $N \times N$ blocks each having dimensions of $3 \times 3$, with the $(\mu,\nu)$-th block of $\bcal{D}$ containing the components of the diffusion tensor $\bm{D}_{\mu\nu }$, and the corresponding block of $\bcal{B}$ being equal to $\bm{B}_{ \mu\nu}$, the decomposition rule for obtaining $\bcal{B}$ can be expressed as $\bcal{B} \cdot {\bcal{B}}^\textsc{t} = \bcal{D} \label{decomp}$. The hydrodynamic tensor $\bm{\Omega}$ is assumed to be given by Rotne-Prager-Yamakawa (RPY) tensor
\begin{equation}
\bm{\Omega}(\bm r^*) =  {\Omega_1{ \boldsymbol \delta} +\Omega_2\frac{\bm r^* \bm r^*}{{r}^{*2}}},
\end{equation}
with
\begin{equation*}
\Omega_1 = \begin{cases} \dfrac{3\sqrt{\pi}}{4} \dfrac{h^*}{r^*}\left({1+\dfrac{2\pi}{3}\dfrac{{h^*}^2}{{r}^{*2}}}\right) \hphantom{{}+{}}& \text{for} \quad r^*\ge2\sqrt{\pi}h^* \\
1- \dfrac{9}{32} \dfrac{r^*}{h^*\sqrt{\pi}} & \text{for} \quad r^*\leq 2\sqrt{\pi}h^* ,
\end{cases}
\end{equation*}
and 
\begin{equation*}
\Omega_2 = \begin{cases} \dfrac{3\sqrt{\pi}}{4} \dfrac{h^*}{r^*} \left({1-\dfrac{2\pi}{3}\dfrac{{h^*}^2}{{r}^{*2}}}\right) \hphantom{{}+{}}& \text{for} \quad r^*\ge2\sqrt{\pi}h^* \\
\dfrac{3}{32} \dfrac{r^*}{h^*\sqrt{\pi}} & \text{for} \quad r^*\leq 2\sqrt{\pi}h^* , 
\end{cases}
\end{equation*}
Here, the hydrodynamic interaction parameter $h^*$ is the dimensionless bead radius in the bead-spring chain model and is defined by $h^* = a/\sqrt{\pi k_BT/k_s}$. 

Since we are interested in the 3D organization of chromatin, we use a number of different static properties to describe the shape of the equilibrium chain. The radius of gyration of the chain, $R_\text{g} \equiv \sqrt{\langle R_\text{g}^2 \rangle}$, where  $\langle R_\text{g}^2 \rangle$ is defined by 
\begin{equation}
\label{rg}
\langle R_\text{g}^2\rangle=\langle\lambda^2_1\rangle+\langle\lambda^2_2\rangle+\langle\lambda^2_3\rangle
\end{equation}
with, $\lambda^2_1$, $\lambda^2_2$, and $\lambda^2_3$ being the eigenvalues of the gyration tensor $\textbf{G}$ (arranged in ascending order), with
\begin{equation}
\label{eq:gy}
\textbf{G} =  \frac{1}{2N_\text{b}^{2}}\sum_{\mu=1}^{N_\text{b}}\sum_{\nu=1}^{N_\text{b}} \textbf{r}_{\mu\nu} \textbf{r}_{\mu\nu}
\end{equation}
Note that, $\textbf{G}$, $\lambda^2_1$, $\lambda^2_2$, and $\lambda^2_3$ are calculated for each trajectory in the simulation before the ensemble averages are evaluated. 
The asymmetry in equilibrium chain shape has been studied previously in terms of various functions defined in terms of the eigenvalues of the gyration tensor~\citep{Kuhn1934,Solc1971,Zifferer1999,Haber2000,Steinhauser2005a,Theodorou1985,Bishop1986}.
Apart from $\lambda^2_1$, $\lambda^2_2$, and $\lambda^2_3$, themselves, we have examined the following \textit{shape functions}: the asphericity ($B$), the acylindricity  ($C$), the degree of prolateness ($S$), and the shape anisotropy ($\kappa^{2}$), as defined in Table~\ref{table2}.

\begin{table*}[t]
	\centering
	\caption{\label{table2} Definitions of shape functions in terms of eigenvalues of the gyration tensor, $\textbf{G}$. Note that, $I_{1} = \lambda^2_1 + \lambda^2_2 + \lambda^2_3$, and $I_{2}= \lambda^2_1 \lambda^2_2 +  \lambda^2_2 \lambda^2_3 +  \lambda^2_3 \lambda^2_1 $, are invariants of $\textbf{G}$.}
	\begin{center}
		\begin{tabular}{lc}
			\hline \hline  Shape function & Definition \\ 
			\hline 
			Asphericity~\citep{Theodorou1985,Soysa2015} &  
			\parbox{6cm}{
				\begin{equation}
				\label{eq:sph}
				B = \langle \lambda^2_{3}\rangle - \frac{1}{2} \left[ \langle \lambda^2_{1}\rangle + \langle \lambda^2_{2}\rangle \right] 
				\end{equation}}  \\
			Acylindricity~\citep{Theodorou1985,Soysa2015} &   
			\parbox{6cm}{
				\begin{equation}
				\label{eq:cyl}
				C =  \langle \lambda^2_{2}\rangle - \langle \lambda^2_{1}\rangle 
				\end{equation}}  \\     
			Degree of prolateness~\citep{Bishop1986,Zifferer1999,Soysa2015} &
			\parbox{9cm}{
				\begin{equation}
				\label{eq:pro1}
				S = \frac{ \left\langle (3\lambda_{1}^2 -  I_{1}) (3\lambda_{2}^2 -  I_{1})(3\lambda_{3}^2 -  I_{1})\right\rangle}{\left\langle \left( I_{1} \right)^{3} \right\rangle}
				\end{equation} }   \\
			Relative shape anisotropy~\citep{Theodorou1985,Bishop1986,Zifferer1999,Soysa2015}    &  
			\parbox{6cm}{
				\begin{equation}
				\label{eq:kappa1}
				\kappa^{2}  = 1 - 3 \frac{\left\langle I_{2} \right\rangle}{\left\langle I_{1}^{2}  \right\rangle}
				\end{equation}}  \\  
			\hline \hline
		\end{tabular}
	\end{center}
\end{table*} 

The stochastic differential equation (Eq.~\ref{eq:sde_eq}) can be solved with a semi-implicit predictor-corrector algorithm developed in ~\citet{PRABHAKAR2004163}, once all the parameters are specified. However, the strength of interaction $\epsilon_{\mu\nu}$ between any two beads $\mu$ and $\nu$ is unknown \textit{a priori}. Since they control the static conformations of a chain, their values will be different depending on whether the gene is in an `ON' or `OFF' state. Ultimately, the contact probability between any two segments on the gene is determined by the values of $\epsilon_{\mu\nu}$ for all pairs on the gene. 
The parameters that need to be specified for us to carry out the simulations are (i) the hydrodynamic interaction parameter $h^*$, (ii) the natural length of Fraenkel spring $r_0^*$, (iii) the SDK potential parameter $\sigma^*$, (iv) the characteristic length scale $l_0$, and (v) the characteristic time scale $\lambda_0$. We are not probing the dynamic properties of chromatin~\cite{prakash2019universal} in the current work, so we chose $h^* =0$. The natural physical length scale in the problem is the diameter of the bead. We assume that chromatin of size 10kb determines the length scale in our model $l_0$, and we coarse grain 10kb chromatin to one bead. The other two length parameters are determined as $\sigma^*=1$ and $r_0^*=1$ such that two neighbouring beads are typically at a distance of the order of $l_0$. All our length results are presented in units of $l_0$.  
The time scale in our problem is given by $\lambda_0 = \zeta/4k_s$. The timestep $\Delta t^* = \Delta t /\lambda_0$ is chosen to be $10^{-3}$. This will decide the time intervals in our simulation. However, since we are only presenting steady-state quantities in this work, all the results are independent of time.

In our model, the distance between the neighbouring beads fluctuates about $r_0$ with the value of order $l_0$, which is the equilibrium length of the spring. For the parameters chosen in this work, $r_0 \pm l_0$ can be greater than $\sigma$. This allows the chain to cross itself in order to explore the whole phase-space faster.
However, this is a result of our choice of parameters values, and we can also choose to have a parameter that makes strand passage more difficult. 

In the section below, we first describe the inverse Brownian dynamics algorithm by which the interaction strengths $\epsilon_{\mu\nu}$ can be estimated given the set of contact probabilities $p_{\mu\nu}$. The issue of converting experimental contact counts to contact probabilities is addressed in section~\ref{sec:c2p}.  

\subsection{\label{sec:ibd}Inverse Brownian dynamics (IBD)}

\begin{figure}[hbt!] 
	\begin{center}
			{\includegraphics*[width=4.0in,height=!]{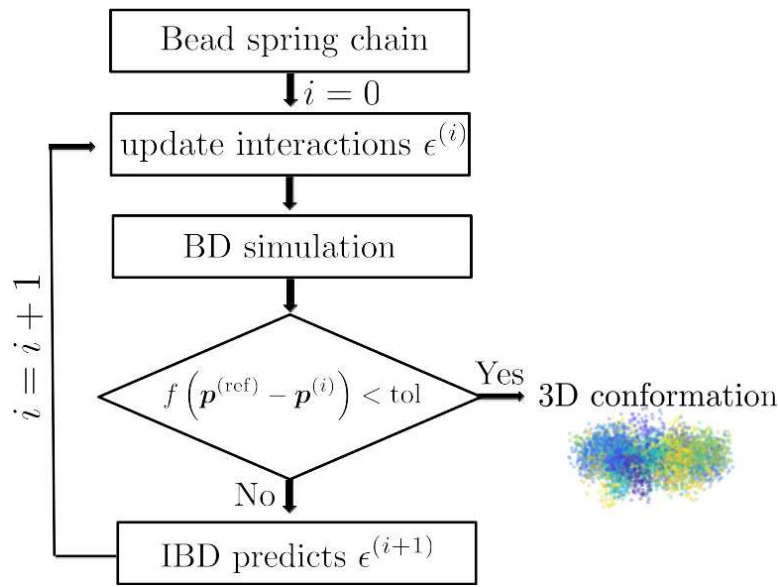}} \\
	\end{center}
	\caption{ Flowchart for the Inverse Brownian Dynamics (IBD) method. Here $\bm p^{\rm (ref)}$ represents the reference contact probability matrix, and $\bm p^{(i)}$ represents the contact probability matrix from simulations at iteration $i$. The interaction strength between beads $\mu$ and $\nu$ is given by $\bm \epsilon_{\mu\nu}$. \label{fig:IBD} }
\end{figure}

In the present investigation, a well-established standard method is utilized to optimise the parameters of a model Hamiltonian, such that it reproduces, as closely as possible, the values of some externally given quantities (e.g.\ from experiment or from other simulations). In the literature, the method is typically referred to as ``Inverse Monte Carlo''~\citep{Lyubartsev1995,Lyubartsev2002,Lyubartsev2015}. It is, however, completely independent of the underlying sampling scheme, as long as the latter produces thermal averages in the canonical ensemble. We prefer to highlight the underlying BD sampling of this study and hence refer to it here as the ``Inverse Brownian Dynamics'' (IBD) method. The method is best explained in general terms. It is assumed that the system is described by a phase space variable $\Gamma$ and a model Hamiltonian $\mathcal{H}(\Gamma)$. Another assumption is that the simulation produces the canonical average of some observable, given by a phase-space function $A(\Gamma)$:
\begin{align}\label{eq:11}
\left\langle A \right\rangle = \dfrac{\int d\Gamma \,A(\Gamma) \,\exp (-\beta\mathcal{H} (\Gamma))}{\int d \Gamma \exp(-\beta \mathcal{H}(\Gamma))} .
\end{align}
Here $\beta = 1/(k_B \textrm{T})$. On the other hand, we have a given ``target'' value $A_t$ (e.g.\ from experiment), which will typically differ from our simulation result. We are now interested in the dependence of the Hamiltonian on some coupling parameter $J$, and we wish to adjust $J$ in order to bring $\left\langle A \right\rangle$ as closely to $A_t$ as possible, within the limitations of the Hamiltonian as such in general, and its dependence on $J$ in particular. In order to do this, it is desirable to obtain information on (i) in which direction $J$ should modified, and (ii) by what amount (at least by order of magnitude). If the change of the coupling constant, $\Delta J$, is small, we can write down a Taylor expansion around the value $J = J_0$ where we performed the simulation:
\begin{align}\label{eq:12}
\left\langle A \right\rangle (J_0 + \Delta J) = \left\langle A \right\rangle(J_0) + \chi \Delta J + O(\Delta J^2),
\end{align}
where the ``generalized susceptibility'' $\chi$ is an abbreviation for the thermodynamic derivative
\begin{align}
\chi = {\dfrac{\partial \left\langle A \right\rangle}{\partial J}\vrule}_{J=J_0} .
\end{align}
The crucial point is now that $\chi$ can be directly sampled in the simulation, by making use of a standard fluctuation relation. Indeed, taking the derivative of Eq.~\ref{eq:11} with respect to $J$, one finds directly
\begin{align}\label{eg:susceptibility}
\chi = \beta \left[\left\langle AB\right\rangle- \left\langle A\right\rangle \left\langle B\right\rangle \right],
\end{align}
where $B$ denotes another phase-space function, which is just the observable conjugate to $J$:
\begin{align}
B(\Gamma) = - \dfrac{\partial\mathcal{H} (\Gamma)}{\partial J }.
\end{align}
In deriving Eq.~\ref{eg:susceptibility}, it is assumed that the phase-space function $A(\Gamma)$ does
not depend on $J$, i.e. $\partial A (\Gamma)/\partial J = 0$. This is the case for most typical applications, and certainly for the present investigation.

The simplest way to do IBD, therefore, consists of (i) neglecting all nonlinear terms in Eq.~\ref{eq:12}, (ii) setting its left hand side equal to $A_t$, (iii) solving for $\Delta J$, and (iv) taking $J_0 + \Delta J$ as a new and improved coupling parameter. The entire process is then repeated with the updated coupling parameter. In other words, Brownian dynamics simulations are carried out again and the difference between the updated  simulation value and the reference value of the observable is compared with the prescribed tolerance and checked to see if convergence has been achieved. If not, the coupling parameters are updated once more until convergence has been achieved. The schematic representation of the IBD algorithm described here is displayed as a Flowchart in Fig.~\ref{fig:IBD}. To avoid overshoots, it is often advisable to not update $J$ by the full increment $\Delta J$ that results from solving the linear equation, but rather only by $\Delta J = \lambda \Delta J$, where $\lambda$ is a damping factor with $0 <\lambda < 1$. The iteration is terminated as soon as $ \lvert \left\langle A \right\rangle \rvert - A_{t}$ does not decrease any more, within some tolerance. One also has to stop as soon as $\chi$ becomes zero, within the statistical resolution of the simulation (this is, however, not a typical situation).

The method may be straightforwardly generalized to the case of several observables $A_m$ and several coupling parameters $J_n$, where the number of observables and the number of couplings may be different. The Taylor expansion then reads
\begin{align}\label{eq:10}
\left\langle A_{m} \right\rangle (\bm J_0 + \Delta \bm{J}) = \left\langle A_{m} \right\rangle(\bm{J}_0) + \sum_{n}\chi_{mn}\, \Delta J_n + O(\Delta \bm J^2),
\end{align}
where the matrix of susceptibilities is evaluated as a cross-correlation matrix:
\begin{align}
\chi_{mn} = \beta \left[\left\langle A_{m}B_{n}\right\rangle- \left\langle A_{m}\right\rangle \left\langle B_{n}\right\rangle \right],
\end{align}
with 
\begin{align}
B_{n}(\Gamma) = - \dfrac{\partial\mathcal{H} (\Gamma)}{\partial J_{n} }.
\end{align}
Typically, the matrix $\chi_{mn}$ will not be invertible (in general, it is not even square!). Therefore, one should treat the linear system of equations via a \textit{singular-value decomposition} (SVD) and find $\Delta J$ via the \textit{pseudo-inverse} (PI). In practice, this means that one updates the couplings only in those directions and by those amounts where one has a clear indication from the data that one should do so, while all other components remain untouched. For details on the concepts of SVD and PI, the reader may refer to~\citet{PresTeukVettFlan92} and~\citet{Fill1998}.

In the present instance, the averages $\left\langle A_{m} \right\rangle$ are the contact probabilities as produced by the simulations, while the target values are the corresponding experimental
values (discussed in greater detail below). The corresponding phase–space functions can be written as indicator functions, which are one in case of a contact and zero otherwise. The coupling parameters that we wish to adjust are the well depths of the SDK attractive interactions, which we allow to be different for each monomer pair. The IBD algorithm discussed here in general terms is described in more detail in section~S1 of the Supporting Material and applied to the specific problem considered here, along with a discussion of the appropriate SVD and PI.    

\section{Results and discussion}

\subsection{\label{sec:val} Validation of the inverse Brownian  dynamics method with a prototype} 
\begin{figure}[hbt!] 
	\begin{center}
		\begin{tabular}{c c}
			{\includegraphics*[width=2.3in,height=!]{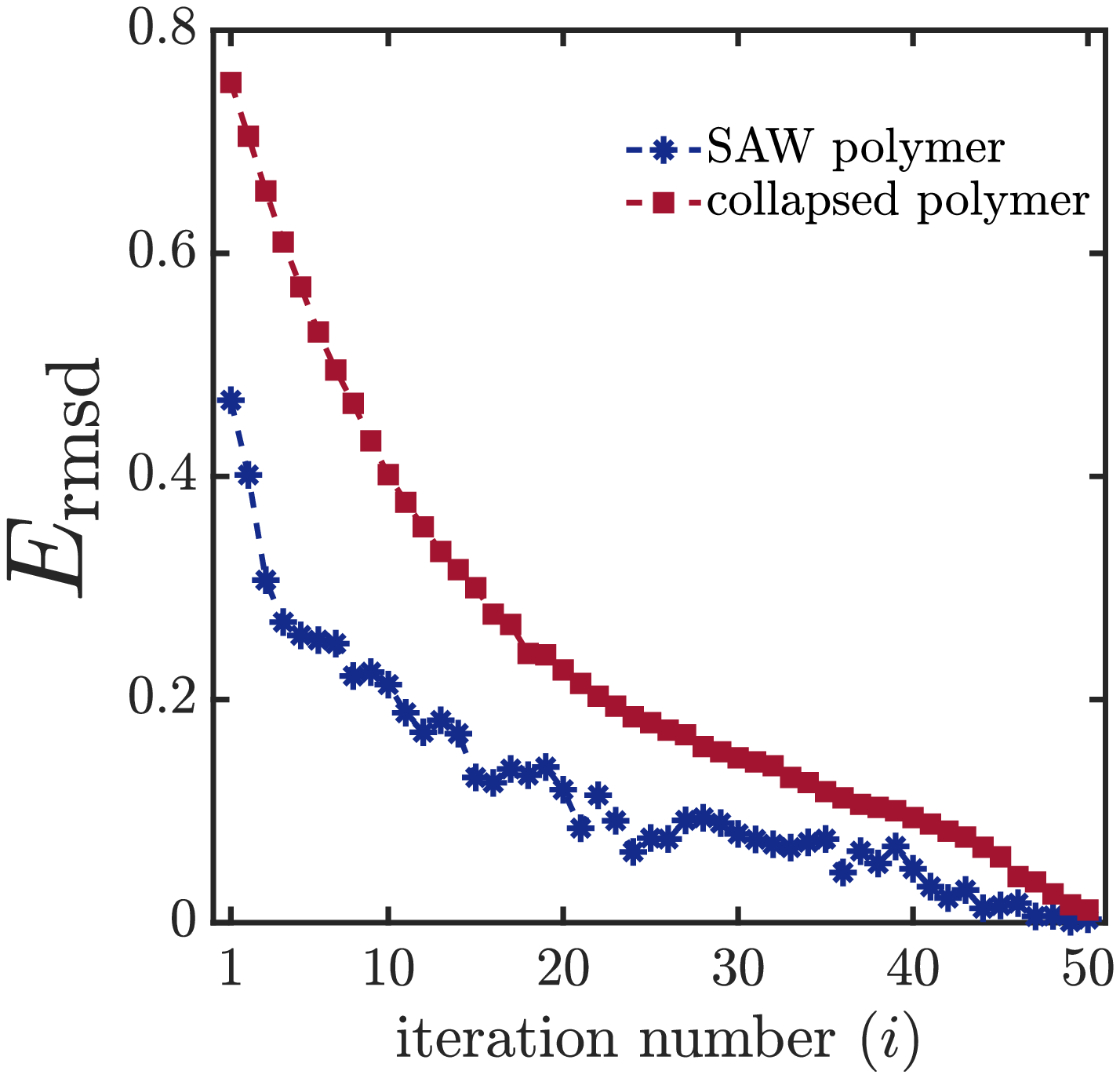}} &
			{\includegraphics*[width=2.7in,height=!]{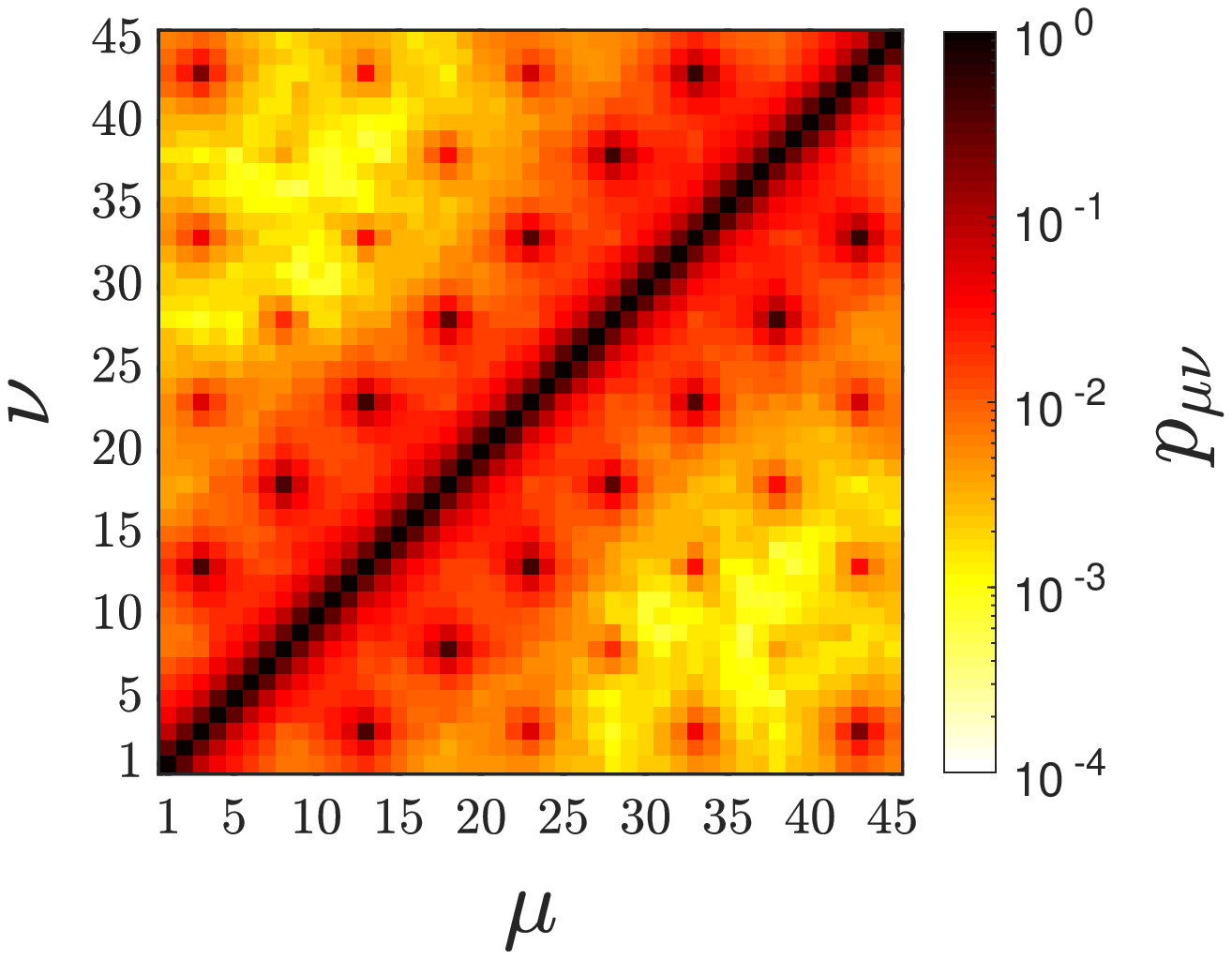}}\\
			(a) & (b)\\
		   	{\includegraphics*[width=2.7in,height=!]{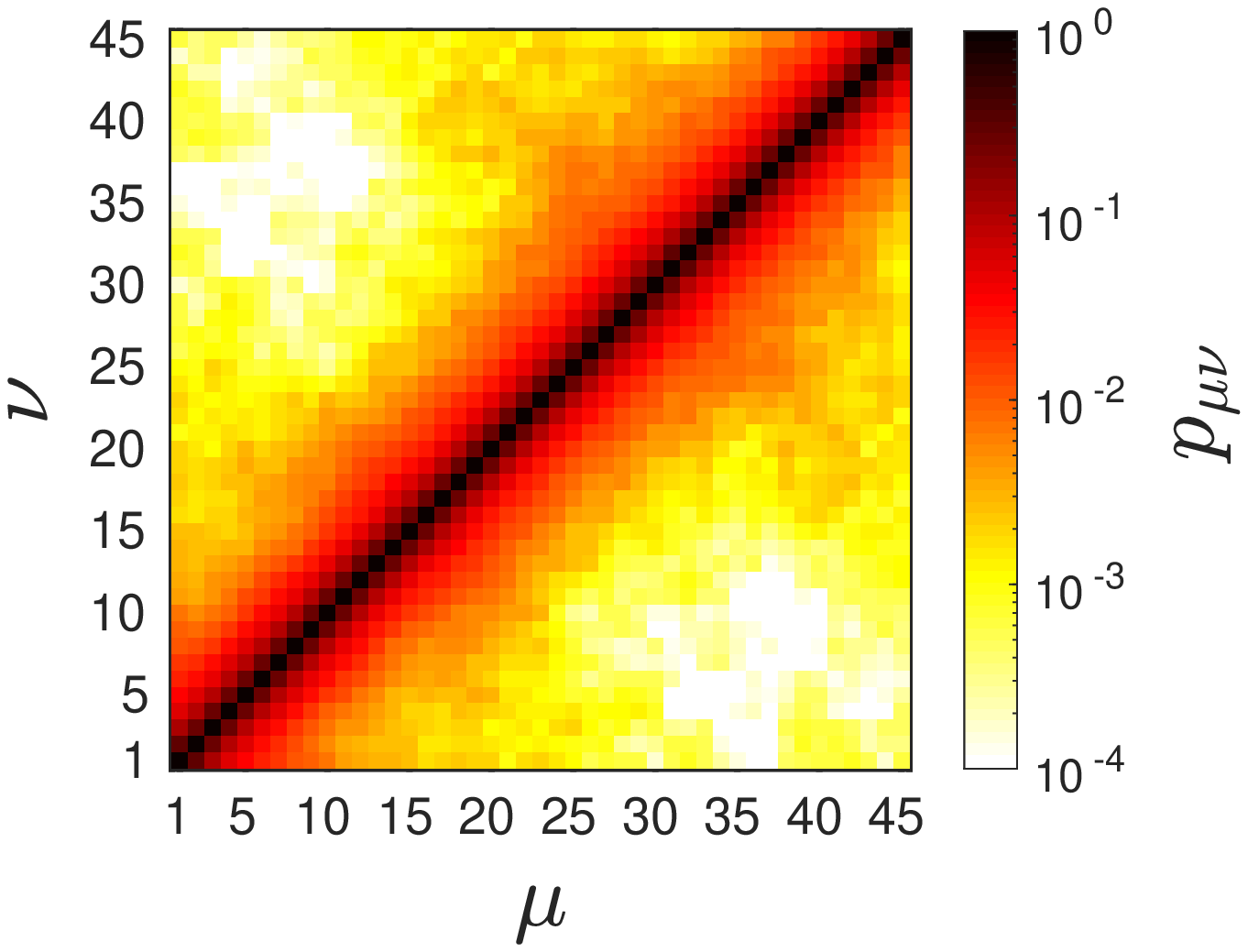}} &
	      	{\includegraphics*[width=2.7in,height=!]{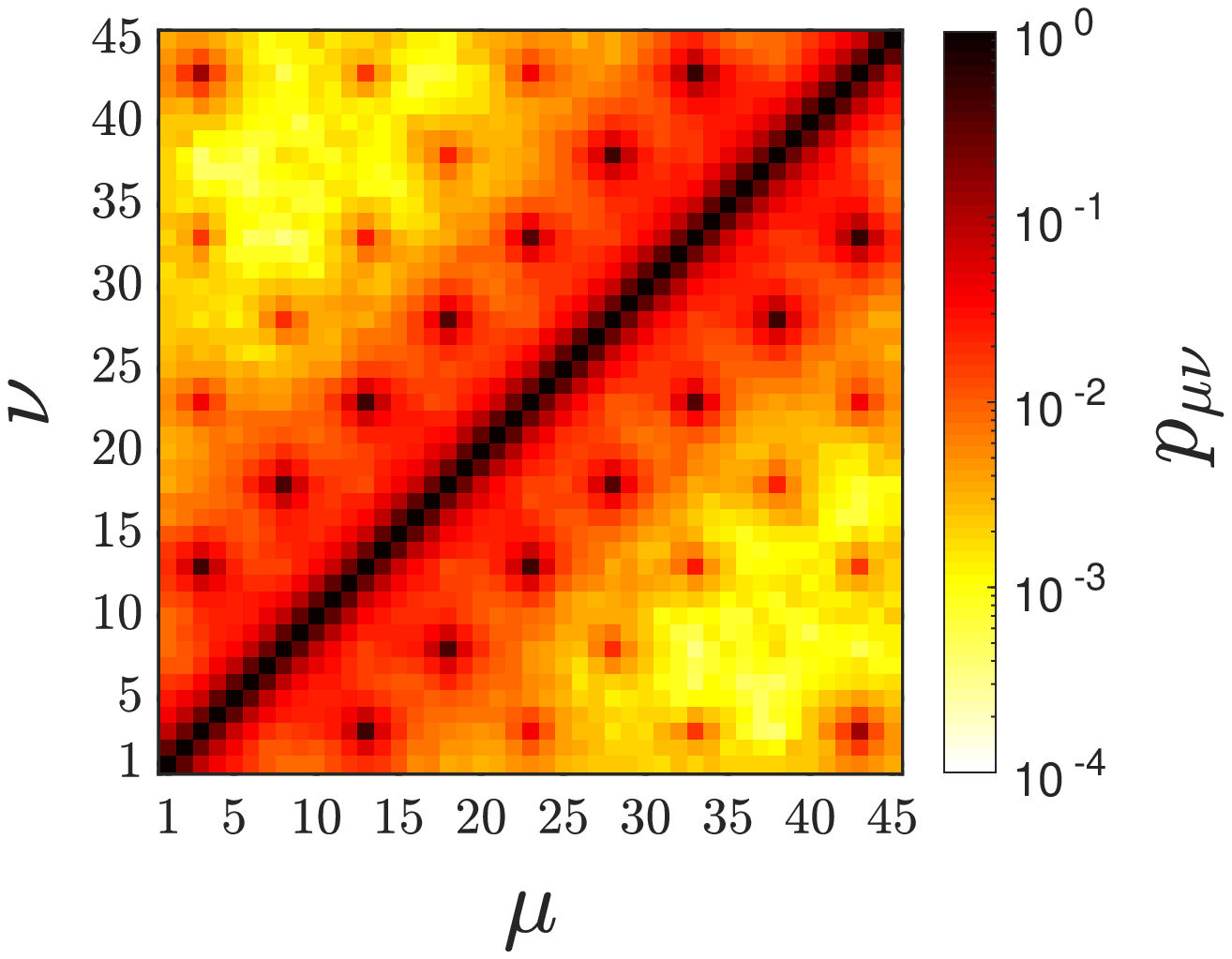}} \\
	        (c) & (d) \\
	        {\includegraphics*[width=2.7in,height=!]{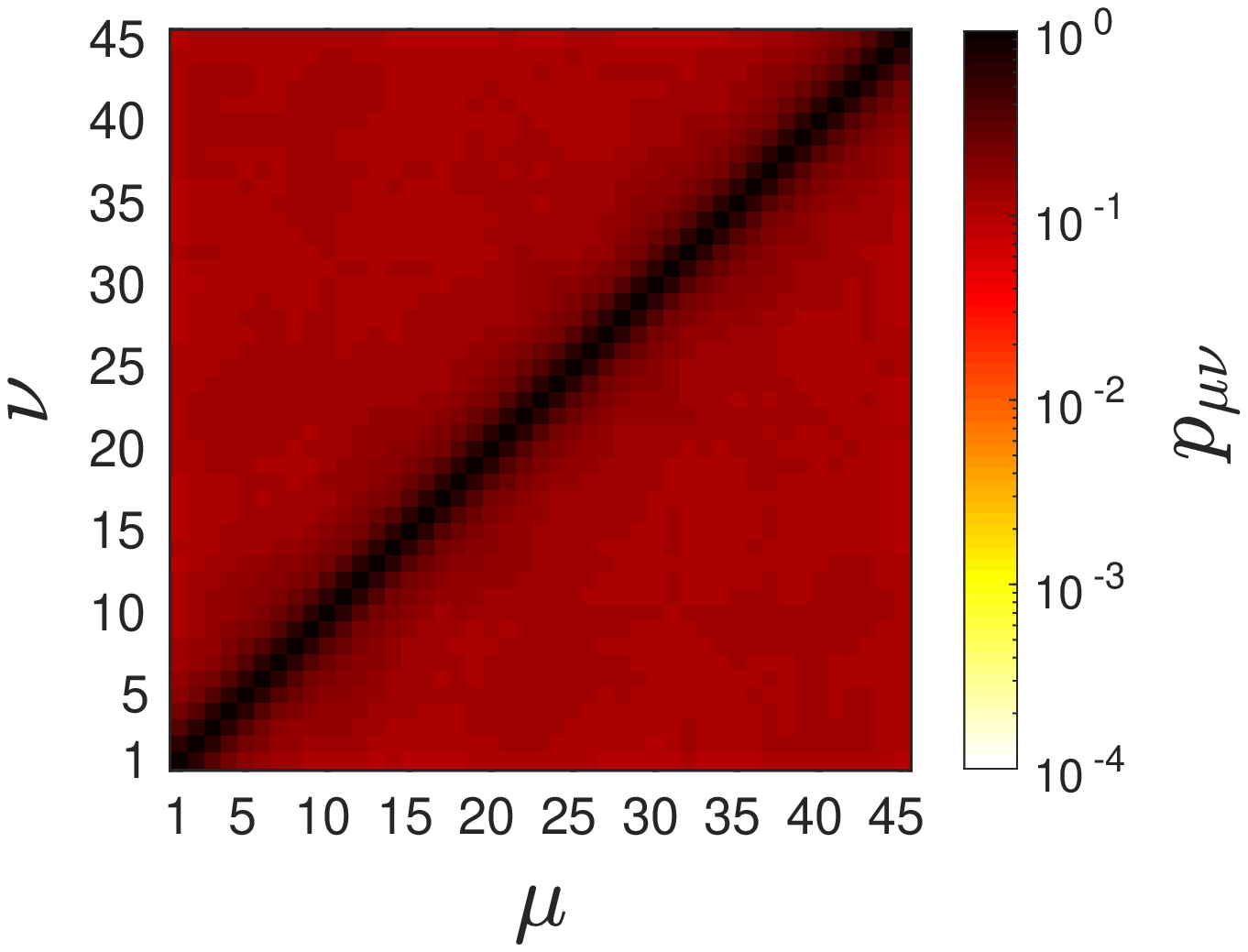}} &
	        {\includegraphics*[width=2.7in,height=!]{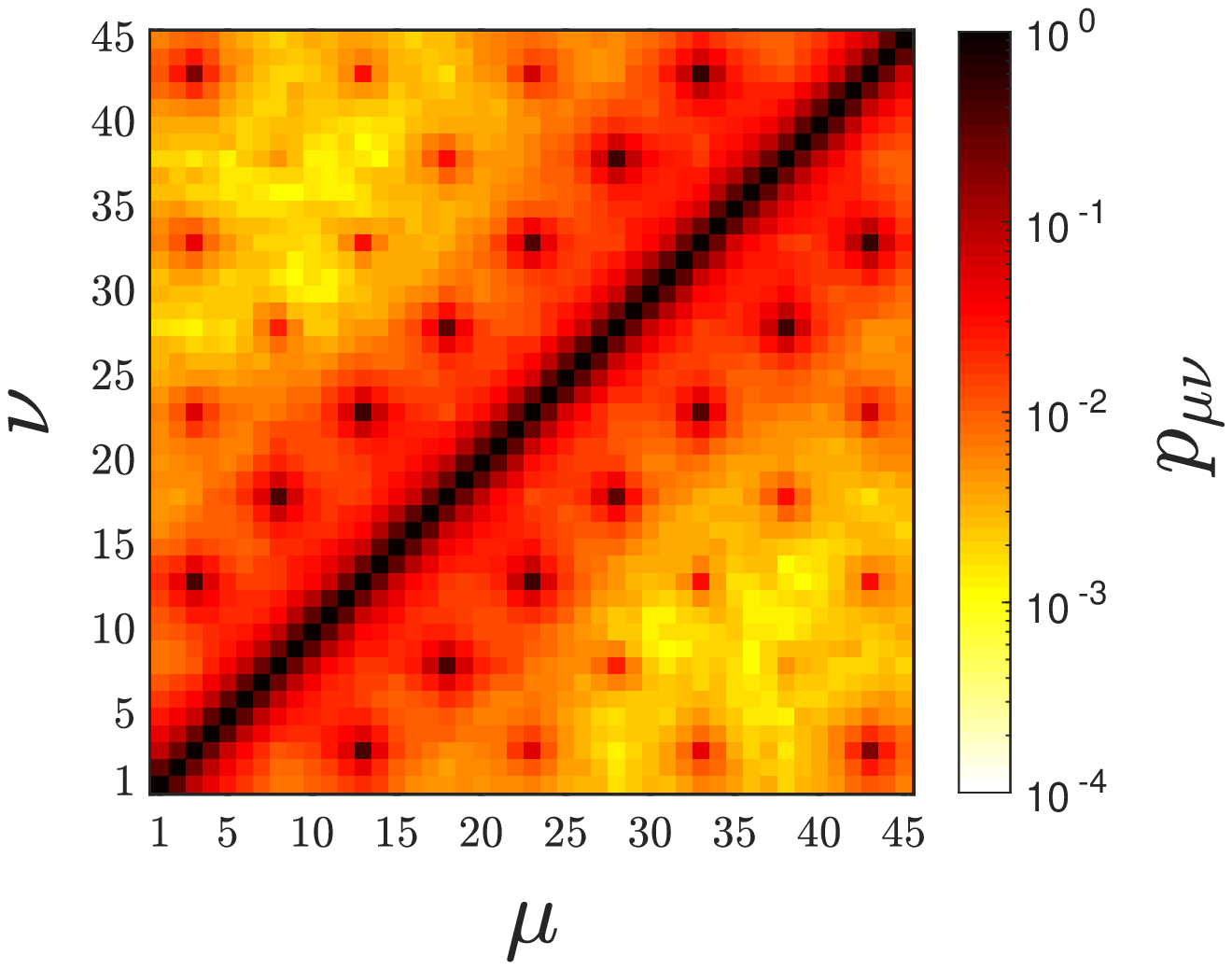}} \\
	        (e) & (f) \\
	    \end{tabular}
	\end{center}
	\caption{Validation of the IBD method with a prototype bead-spring chain with 45 beads. (a) Root-mean-square deviation $E_{\rm rmsd}$ (Eq.~\ref{eq-rmsd}) as a function of iteration number showing convergence of the IBD method. (b) Reference contact probability matrix. 
		Two different initial states have been considered for testing IBD convergence: (c) initial contact probability for the self-avoiding walk (SAW) where no bead-pairs have attractive interaction and (d) recovered contact probability matrix through IBD starting from the SAW state. Similarly (e) initial contact probability for the collapsed state where all the bead-pairs have attractive interaction, $\epsilon = 1$ and (f) recovered contact probability matrix through IBD starting with the collapsed state. The abscissa and ordinate represent the bead number along the polymer chain. The colour represents the contact probability between the beads $\mu$ and $\nu$ (see colour bar).
		\label{fig:ibd_45}}
\end{figure}

To validate the IBD method, a prototype of a chromatin-like polymer chain with artificially set interaction strengths ($\epsilon_{\mu\nu}$) was constructed. The data from this simulated chain was used to test the IBD algorithm, as described below. The IBD algorithm was validated  for chains of length 10, 25 and 45 beads. Here we discuss the 45 bead chain case as a prototype. 
A few bead-pairs $(\mu\nu)$ were connected arbitrarily with a prescribed value of the well-depth $\epsilon_{\mu\nu}^{\rm (ref)}$ of the SDK potential. The non-zero reference interaction strengths for the connected bead-pairs $ \epsilon_{\mu\nu}^{\rm (ref)}$ are shown in Table~\ref{table:val}; the remaining pairs were considered to have no attractive interaction ($\epsilon_{\mu\nu}^{\rm (ref)}=0$). 
The beads-spring chain was simulated until it reached equilibrium, which was quantified by computing $R_g$ as a function of time. A stationary state was observed to be reached after eight Rouse relaxation times~\citep{Marrucci1989}. However, equilibration was continued for a further fifteen Rouse relaxation times. After equilibration, an ensemble of $10^5$ polymer configurations was collected from 100 independent trajectories, from each of which $10^3$ samples were taken at intervals of $10^3$ dimensionless time steps, which correspond to roughly 2 to 3 Rouse relaxation times. From this ensemble, the contact probability $ p_{\mu\nu}^{\rm (ref)} =\langle{ \hat{p}_{\mu\nu} \rangle}$ for each bead pair in the chain was computed. Here $\hat{p}_{\mu\nu}$ is an indicator function which is equal to $1$ or $0$ depending upon whether the $\mu^{\rm th}$ and $\nu^{\rm th}$ beads are within the cut-off distance of SDK potential ($r_{\mu\nu}^* \leq r_{\rm c}^{*}$) or not ($r_{\mu\nu}^* > r_{\rm c}^{*}$). The reference contact probabilities $p_{\mu\nu}^{\rm (ref)}$, determined in this manner, are shown in Fig.~\ref{fig:ibd_45}(b). In the present instance while $p_{\mu\nu}^{\rm (ref)}$ has been constructed by simulating the bead-spring chain for the given values of $\epsilon_{\mu\nu}^{\rm (ref)}$, in general it refers to the experimental contact probabilities.

\begin{table}[h!]
	\setlength{\tabcolsep}{14pt}
	\centering
	\caption{\label{table:val} Interaction strengths $\epsilon_{\mu\nu}^{\rm}$ and contact probabilities $p_{\mu\nu}^{\rm}$ for selected bead pairs $(\mu, \nu)$ in a bead-spring chain with 45 beads. Values of these variables recovered using IBD are compared with those of the reference polymer chain, along with the percentage error between the reference and recovered values. Initial $\epsilon_{\mu\nu}$ values for all the bead-pairs were chosen to be 0 for the self-avoiding walk polymer while $\epsilon_{\mu\nu}=1$ for all the bead-pairs in the collapsed polymer.}
	\label{tab:val}
	\begin{tabular}{|c|ccc|cccc|} 
		\hline
		\multicolumn{8}{|c|}{Initial state: self-avoiding walk polymer}                                                                                                            \\ \hline
		\multirow{2}{*}{bead-pair} & \multicolumn{3}{c|}{interaction strength, $\epsilon_{\mu\nu}$} & \multicolumn{4}{c|}{contact probability, $p_{\mu\nu}$} \\ \cline{2-8} 
		& reference           & recovered           & \% error           & initial & reference         & recovered         & \% error        \\ \hline
		3-13                       & 7.00                   & 6.70                & 4.29       &  0.0033      & 0.44              & 0.46              & 4.55            \\
		13-23                      & 7.00                  & 7.28                & 4.00        & 0.0036      & 0.51              & 0.49              & 3.92            \\
		23-33                      & 7.00                   & 7.08                & 1.14       & 0.0057       & 0.39              & 0.37              & 5.13            \\
		33-43                      & 7.00                   & 7.35                & 5.00       & 0.0041       & 0.62              & 0.59              & 4.84            \\
		8-18                       & 7.00                   & 6.94                & 0.86       & 0.0056       & 0.47              & 0.47              & 0.00            \\
		18-28                      & 7.00                   & 6.89                & 1.57       & 0.0052       & 0.31              & 0.32              & 3.23            \\
		28-38                      & 7.00                   & 7.16                & 2.29       & 0.0071       & 0.55              & 0.53              & 3.64            \\
		3-43                       & 7.00                   & 7.18                & 2.57       &  0.0002      & 0.22              & 0.22              & 0.00  
		\\ \hline    
		\multicolumn{8}{|c|}{Initial state: collapsed polymer}                                                                                                            \\ \hline
		\multirow{2}{*}{bead-pair} & \multicolumn{3}{c|}{interaction strength, $\epsilon_{\mu\nu}$} & \multicolumn{4}{c|}{contact probability, $p_{\mu\nu}$} \\ \cline{2-8} 
		& reference           & recovered           & \% error           & initial & reference         & recovered         & \% error        \\ \hline
		3-13                       & 7.00                   &  6.67               & 4.71      &    0.139    & 0.44              &  0.44             &   0.00          \\
		13-23                      & 7.00                  &6.99                 &  0.14       &  0.141     & 0.51              &  0.52            &  1.96           \\
		23-33                      & 7.00                   & 6.75                &   3.57     &0.133        & 0.39              &    0.38           & 2.56            \\
		33-43                      & 7.00                   &  7.19               & 2.71      &  0.136      & 0.62              &   0.59           &  4.84          \\
		8-18                       & 7.00                   &  7.22              &  3.14      & 0.132       & 0.47              &      0.45         &  4.26           \\
		18-28                      & 7.00                   &  6.77               &  3.29     &  0.135      & 0.31              &      0.3        &  3.23          \\
		28-38                      & 7.00                   &  6.89               & 1.57       &   0.133     & 0.55              &     0.55         &  0.00           \\
		3-43                       & 7.00                   &   7.11              &  1.57      & 0.067       & 0.22              &    0.22           &  0.00
		\\ \hline     
	\end{tabular}
\end{table}

The IBD method was then applied to recover the reference contact probabilities $p_{\mu\nu}^{\rm (ref)}$ starting with an initial guess of a self-avoiding walk where $\epsilon_{\mu\nu}^{(0)} = 0$, i.e., all the interaction strengths are set equal to zero. The contact probability for the initial state of self-avoiding walk is shown in Fig.~\ref{fig:ibd_45}(c). As illustrated in Fig.~\ref{fig:IBD}, at each iteration step $i$, Brownian dynamics was performed for the given $\epsilon_{\mu\nu}^{(i)}$ and an ensemble of $10^5$ conformations were collected.  To quantify the difference between contact probabilities computed from simulation at iteration $i$ ($p_{\mu\nu}^{(i)}$) and reference contact probabilities ($p_{\mu\nu}^{(\rm ref)}$), the root mean-squared deviation ${E_{\rm{rmsd}}^{(i)}}$ was calculated
\begin{align}\label{eq-rmsd}
{E_{\rm{rmsd}}^{(i)}} = \sqrt{\frac{2}{N(N-1)}\sum_{1\le \mu< \nu \le N} \left( p_{\mu \nu}^{(i)}- p_{\mu \nu}^{\rm (ref)}\right)^2 }
\end{align}
at each iteration. The error criteria ${E_{\rm{rmsd}}^{(i)}}$ has been used previously in \citet{Meluzzi2013},  and is adopted here. At each iteration $i$, if the ${E_{\rm{rmsd}}^{(i)}}$ value  is greater than the preset tolerance limit (tol),  the interaction strength parameters $\epsilon_{\mu\nu}^{(i+1)}$ for the next iteration were calculated as given in Eq.~S11 (see Supporting Material). To avoid the overshoot in interaction strength $\epsilon_{\mu\nu}^{ (i+1)}$, the range of $ \epsilon_{\mu\nu}^{(i+1)}$ was constrained  to [0, 10]. 
For the investigated polymer chain with $45$ beads, the IBD algorithm converges (${E_{\rm{rmsd}}^{(i)}} <\rm tol$) in approximately 50 iterations and  $p_{\mu \nu}^{\rm ref}$ was recovered.
The error $E_{\rm rmsd}$ for each iteration is shown in Fig.~\ref{fig:ibd_45}(a) while the recovered contact probability matrix is shown in Fig.~\ref{fig:ibd_45}(d). 
The recovered contact probability values along with the optimized interaction strengths $\epsilon_{\mu\nu}$ are shown in Table~\ref{table:val}. The error in the recovered contact probabilities and interaction strengths is less than $5$\%, proving the reliability of the IBD method. The largest contact probabilities are for those bead-pairs for which values of the interaction strength were chosen a priori, as given in Table~\ref{tab:val}. However, the existence of these interactions leads to the existence of contact probabilities $p_{\mu\nu}$ between all bead-pairs $\mu$ and $\nu$. The IBD algorithm was applied to not just the specified bead-pairs but to recover all contact probabilities $p_{\mu\nu}$, for all possible pairs. The errors are given in Table~\ref{tab:val} only for the specified values since they are the largest. 
To check the robustness of the IBD algorithm, the same reference contact probability of the prototype was recovered from an entirely different initial configuration of a collapsed chain where $\epsilon_{\mu\nu}^{(0)} = 1$ for all the bead pairs $\mu$ and $\nu$. The initial contact probability matrix of the collapsed chain is shown in Fig.~\ref{fig:ibd_45}(e) and the recovered contact probability matrix starting from the collapsed chain is shown in Fig.~\ref{fig:ibd_45}(f). The recovered contact probability values along with the optimized interaction strengths $\epsilon_{\mu\nu}$ for a few bead-pairs are shown in Table~\ref{table:val}. Thus, even starting from a very different configuration, the IBD algorithm converges to the target contact probability matrix, establishing the power of the method. For the sake of completeness, the difference between the reference and recovered contact probability matrices is presented in section~S2 of the Supporting Material, along with a discussion of the pathways by which the polymer chain converges from different initial configurations (swollen or collapsed) to the final reference state. Having validated the IBD algorithm, the next section applies this technique to experimentally obtained contact probabilities of a chromatin, on the length scale of a gene.

\subsection{\label{sec:cg} The coarse-graining procedure  }
\begin{figure}[hbt!] 
	\begin{center}
		\begin{tabular}{c c}
			{\includegraphics*[width=2.7in,height=!]{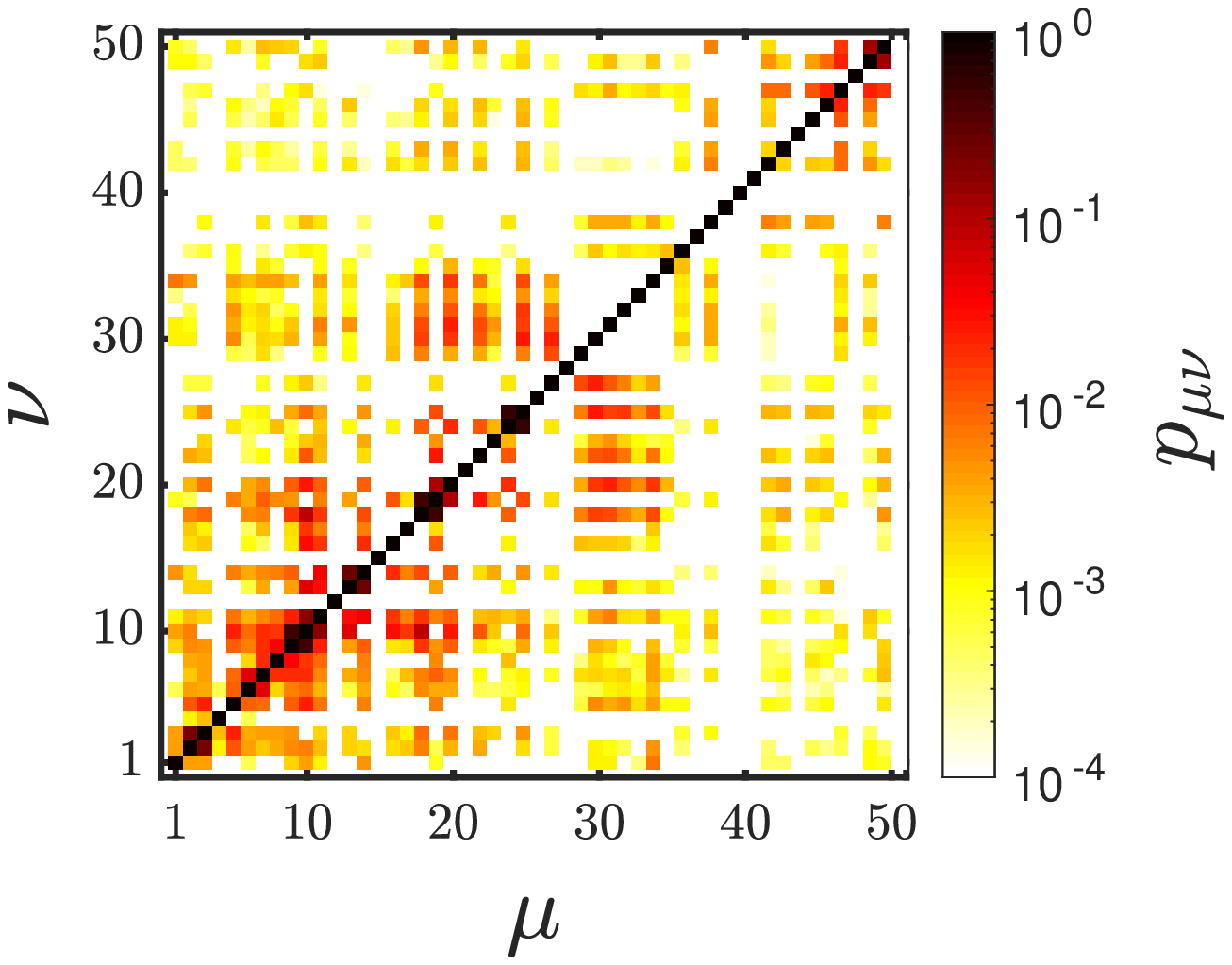}} &
			{\includegraphics*[width=2.7in,height=!]{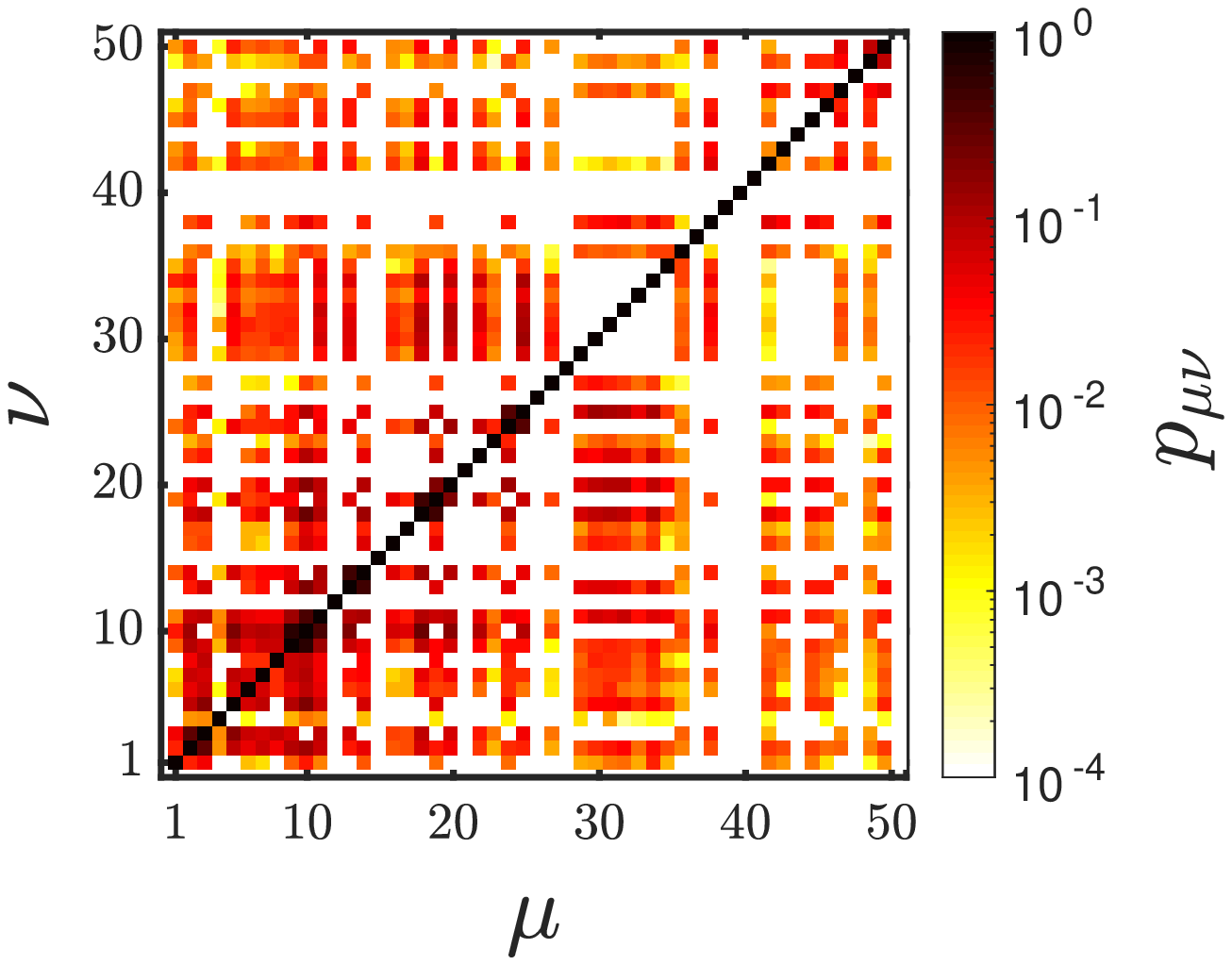}}\\
			(a) K$562$ (ON) - reference CP & (b)GM$12878$ (OFF) - reference CP \\
			\\
			{\includegraphics*[width=2.7in,height=!]{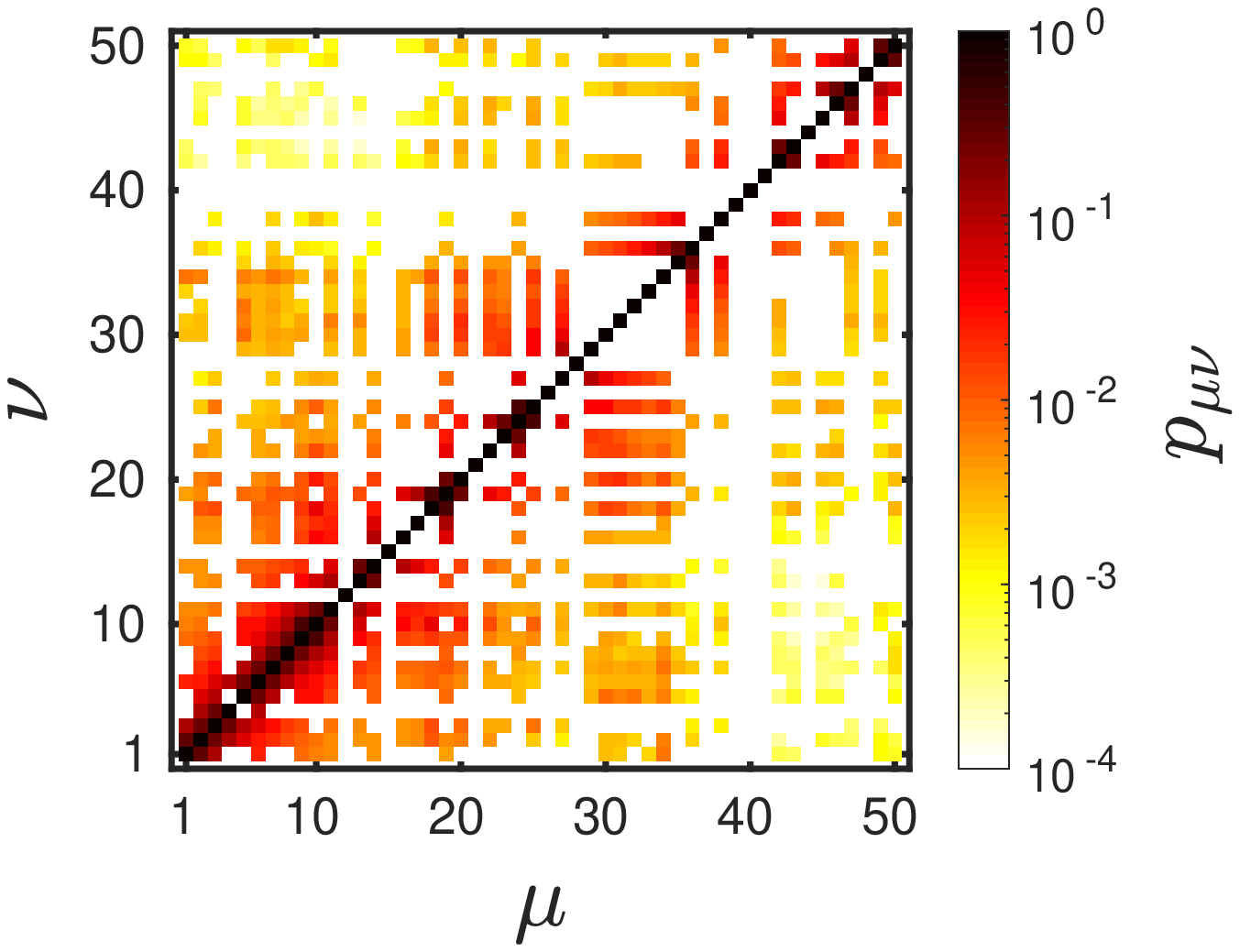}} &
			{\includegraphics*[width=2.7in,height=!]{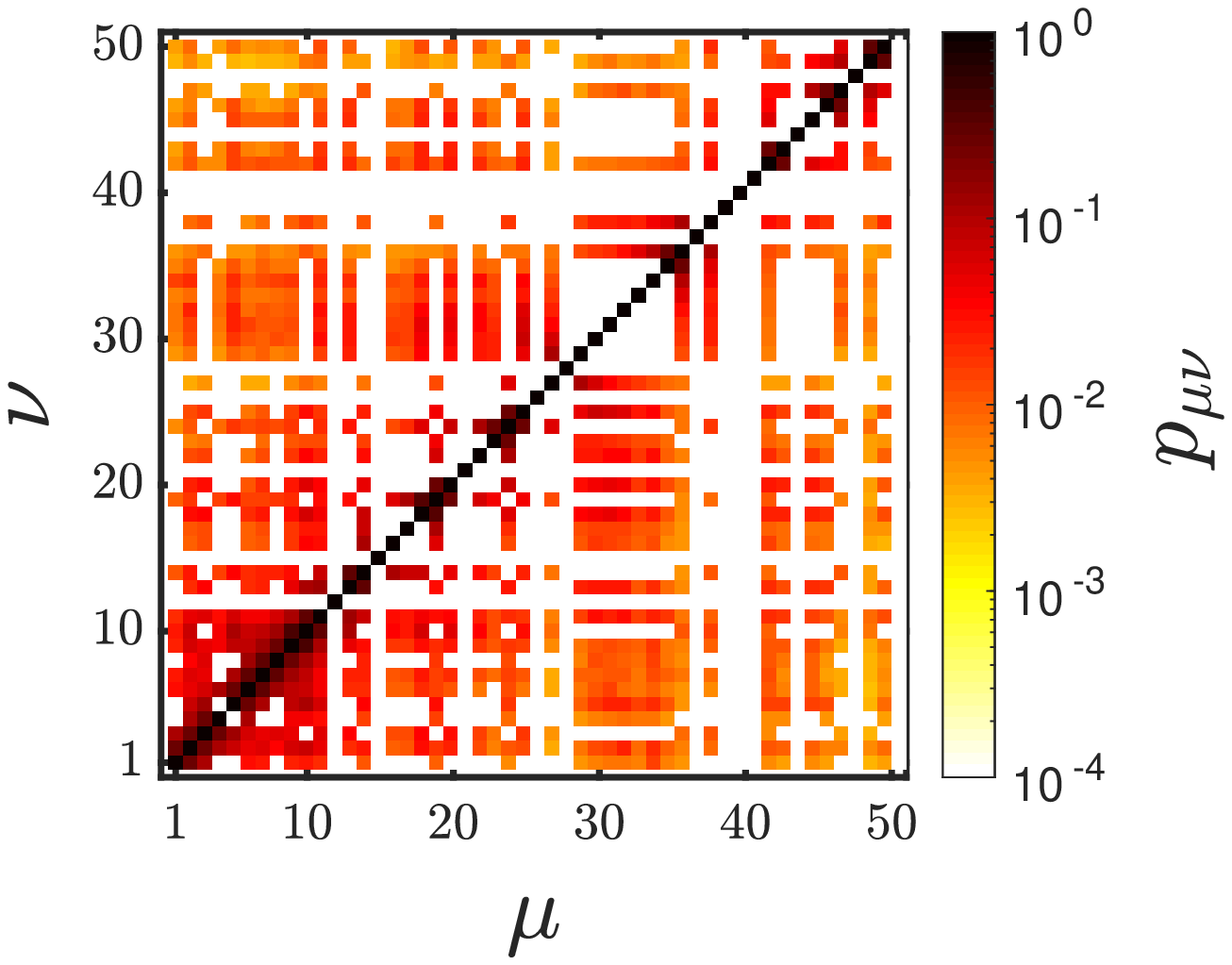}}\\
			(c)  K$562$ (ON) - recovered CP & (d) GM$12878$ (OFF) - recovered CP\\
			\\
			{\includegraphics*[width=2.7in,height=!]{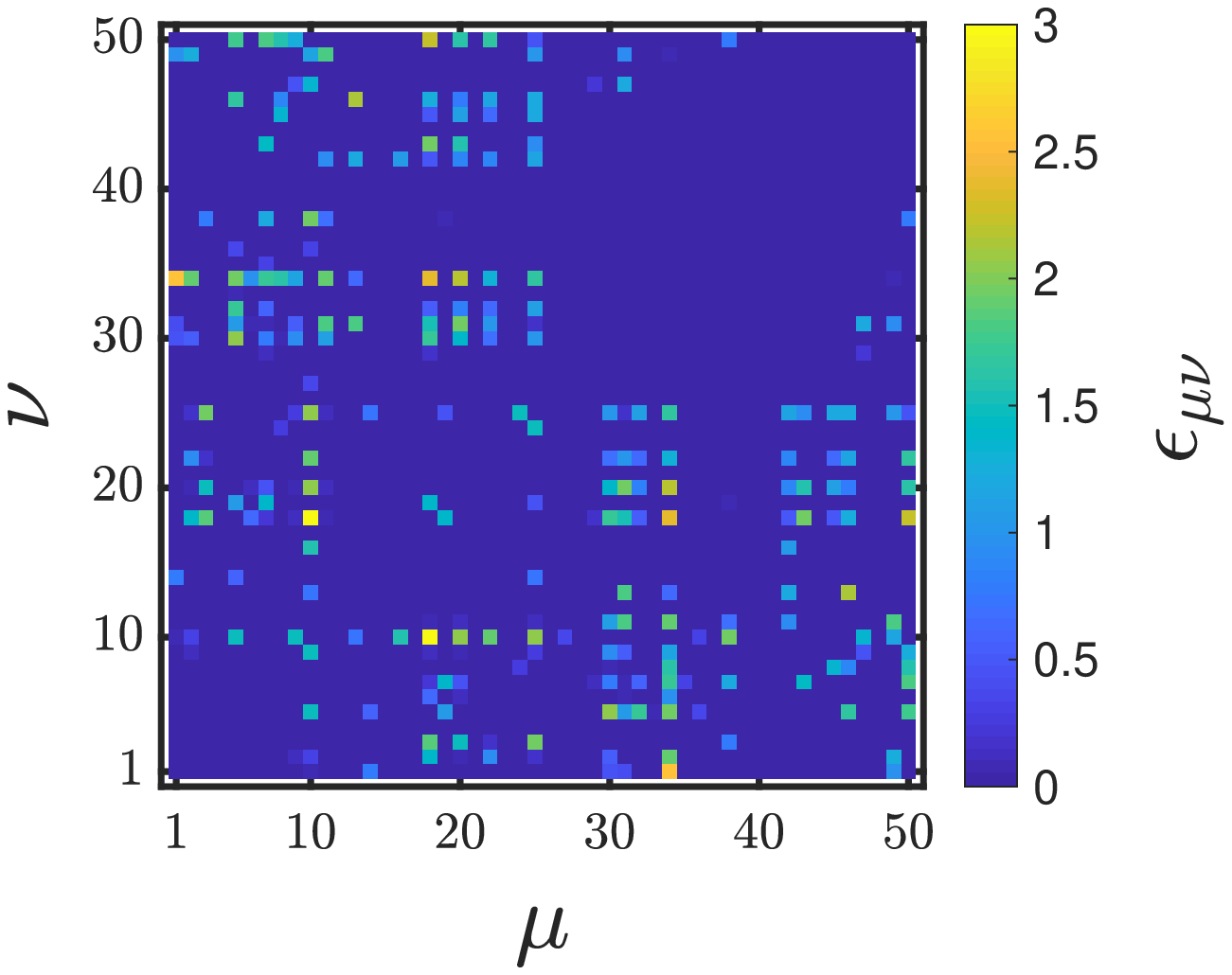}} &
			{\includegraphics*[width=2.7in,height=!]{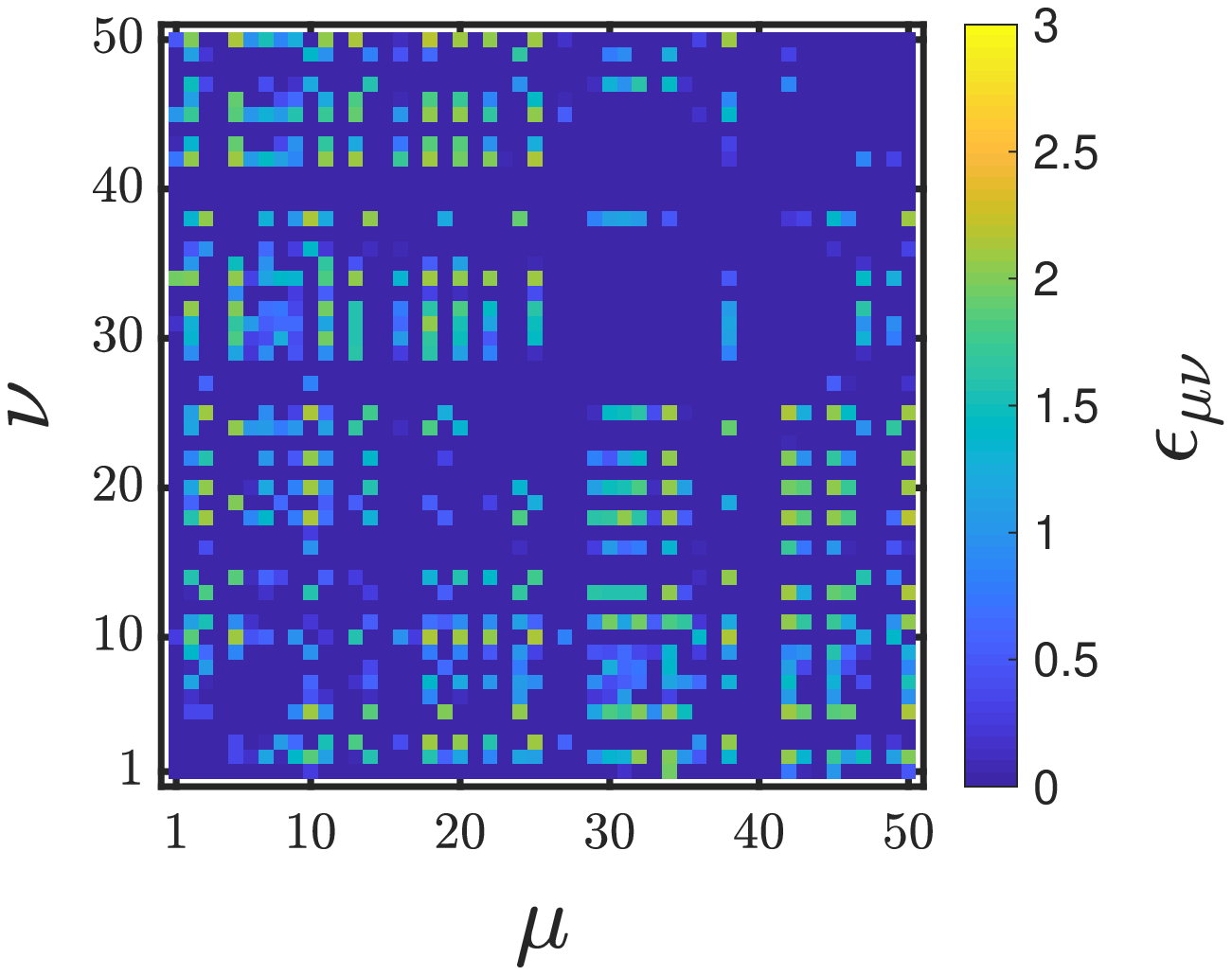}} \\
			(e) K$562$ (ON) - $\epsilon_{\mu\nu}$ & (f) GM$12878$ (OFF) - $\epsilon_{\mu\nu}$\\
		\end{tabular}
	\end{center}
	\caption{Comparison of the reference normalized contact probabilities ((a) and (b)) with the recovered contact probabilities ((c) and (d)), obtained with the IBD method for K$562$ and GM$12878$, respectively, at $N_f = 0$. The value of interaction strength parameter $\epsilon_{\mu\nu}$ for (e) K562 (ON state) and (f) GM12878 (OFF state) cell lines, respectively, at the converged state.
		\label{fig:rec_alpha} }
\end{figure}

To study the 3D organization of a gene region, the $\alpha$-globin gene locus (ENCODE region ENm008) is chosen for which~\citet{Bau2011a} have experimentally determined the contact counts using the 5C technique. This is a 500~kbp long region on human chromosome 16 containing the $\alpha$-globin gene and a few other genes like LUC7L. Since 5C data does not interrogate the contact counts between all feasible 10 kbp segment pairs, many elements in the heat map have no information. This is in contrast with typical Hi-C experiments where information on all possible contact pairs are obtained. In principle, this method can be applied to Hi-C data; however, in this instance, we chose the 5C data since it has sufficiently good resolution.

For simulation purpose, the $\alpha$-globin locus is coarse-grained to a bead-spring chain of $50$ beads. That is, the experimental 5C data (contact count matrix of size $70\times70$) for the EMn008 region was converted to a contact count matrix of size $50\times50$.  The coarse graining procedure is as follows: 500kb of the gene locus was divided into 50 beads, each comprising 10 kb equal-sized fragments. The midpoint of each restriction fragment was located and was assigned to the corresponding bead in the coarse-grained polymer. 
There are cases where two or more restriction fragments (each of size less than 10kbp) get mapped to the same bead. For example, consider restriction fragments $r_1$ and $r_2$ being mapped on to a single coarse-grained bead $\mu$, and fragments $r_3$ and $r_4$ being mapped on to another bead $\nu$. The contact counts of the coarse-grained bead-pair $C_{\mu\nu}$ can then be computed in at least three different ways, namely \textit{independent}, \textit{dependent} and \textit{average} coarse-graining procedures, as described below. 
\begin{itemize}
	\item \textit{Independent coarse graining}: Take the sum of all contact counts for the four restriction fragment combinations ($C_{\mu\nu} = C_{r_1 r_3}+C_{r_1r_4}+C_{r_2r_3}+C_{r_2r_4}$) --- i.e., assume that all contacts occur independently of each other, in other words not more than one of the contact pairs occurs in the same cell.
	
	\item \textit{Dependent coarse graining}: Take the maximum contact count amongst all the four restriction fragment combinations ($C_{\mu\nu} = \max\{C_{r_1r_3},C_{r_1r_4},C_{r_2r_3},C_{r_2r_4}\}$). This assumes that whenever the pairs having small contact counts are in contact, the pair with the largest contact count is also in contact. These are the two extreme cases and the reality could be somewhere in between. 
		
	\item  \textit{Average coarse graining}: The third option is then to choose some such intermediate value. Here, we use the approximation that the coarse grained contact count is equal to the average of the two extreme contact counts mentioned earlier, namely $C_{\mu\nu} =\frac{1}{2}[ (C_{r_1r_3}+C_{r_1r_4}+C_{r_2r_3}+C_{r_2r_4})+\max\{C_{r_1r_3},C_{r_1r_4},C_{r_2r_3},C_{r_2r_4}\}]$. 
	
\end{itemize}

\subsection{\label{sec:c2p} Conversion of contact counts to contact probabilities: the normalization problem}
The contact counts obtained from the Chromosome Conformation Capture experiments are not normalized. That is, the contact count values can vary from experiment to experiment and total number of contacts are not quantified. This data cannot be compared across cell lines or across different experimental sets. 
While several normalization techniques exist, the ICE method is one of the more widely used techniques, where through an iterative process biases are removed and  equal ``visibility'' are provided to each bins/segments of the polymer. The resulting contact count matrix is a normalized matrix where $\sum_{\mu} C_{\mu\nu}=1$.
While the existing normalization techniques help in removing biases, they still only give relative contact probabilities and not the absolute values.
To accurately predict the distance between any two segments in chromatin, it is essential to know their absolute contact probabilities.
Since the total number of genome equivalent (number of cells) cannot be estimated in a chromosome conformation capture experiment, the calculation of absolute contact probability from the contact count is highly challenging. A simple technique to normalize these counts is described here. The contact count matrix can be normalized by imposing the following constraint, namely, that the sum of times any segment pairs $(\mu ,\nu)$ are in contact ($C_{\mu\nu}^{\rm c}$) and the number of times they are not in contact ($C_{\mu\nu}^{\rm nc}$) must be equal to the total number of samples $N_s$. This is true for all bead-pairs i.e. $C_{\mu\nu}^{\rm c}+C_{\mu\nu}^{\rm nc}=N_{s}$, for all $\mu$ $\nu$. Since only $C_{\mu\nu}^{\rm c}$ is known, two limiting values of $N_s$ are estimated using the following scenarios. In one scenario, it is assumed that for the segment pairs $(\mu ,\nu)$ which has the largest contact count in the matrix, $\mu$ and $\nu$ are always in contact in all cells. In other words $C_{\mu\nu}^{\rm nc}=0$; in this case $N_{s}$ is simply equal to the largest element of the contact count matrix. Since this is the smallest value of $N_{s}$ possible, it is denoted by $(C_{\mu\nu}^c)_{\rm max}=N_{\rm min}$. 
The other scenario estimates the sample size from the row $\mu$ for which the sum over all contact counts is the largest i.e., $N_{s}=$ maximum of $(\sum_\nu C_{\mu\nu}^{\rm c})$. This assumes that $\mu$ is always in contact with only one other segment in a cell and there is no situation when it is not in contact with any segment. This case is denoted as $N_{\rm max}$.
However, in a real system, there might be situations where segment $\mu$ is not in contact with any of the remaining segments. In such a case, $N_s$ could be greater than $N_{\rm max}$. We have investigated this question in the context of simulations, where we know the exact ensemble size, and can normalize the contact count matrix with the exact ensemble size, i.e., $N_s$. From this analysis, it was observed that there are very few samples where the bead $\mu$ is not in contact with any of the remaining beads.
It supports our hypothesis that $N_{\rm max}$ could be considered to be the upper limit in estimating the ensemble size $N_s$.
Since the precise value of  $N_{s}$ is not known in experiments, $N_s$ is varied as a parameter from $N_{\rm min}$ to $N_{\rm max}$. To systematically vary $N_s$, for convenience, a parameter $N_f$ is defined,
\begin{align}\label{eq:N_f}
N_f = \frac{N_s - N_{\textrm{min}}}{N_{\textrm{max}}-N_{\textrm{min}}}
\end{align}
in the range of $[0, 1]$. Clearly, $N_f = 0$ implies $N_s = N_{\rm min}$, which is the lower bound for $N_s$ and $N_f = 1$ implies $N_s = N_{\rm max}$, which is the upper bound. 
The contact probabilities at various $N_f$ values are calculated as $p_{\mu\nu} =(C_{\mu\nu}^c/N_{\textrm{s}})$ where $N_s= N_{\textrm{min}}+ N_f (N_{\textrm{max}}-N_{\textrm{min}})$.

For several values of $N_f$, the contact count matrices are normalized and IBD is carried out to obtain the optimal interaction strengths between the bead-pairs. Fig.~\ref{fig:rec_alpha}(a) and \ref{fig:rec_alpha}(b) show the normalized contact probabilities at $N_f=0$ for cell lines K562 (ON state) and GM12878 (OFF state), respectively (reference contact probabilities), when they are coarse-grained to 50 segments of length 10 kbp each, as per the procedure described above and the corresponding recovered contact probability matrices for both the cell lines from simulation are shown in Fig.~\ref{fig:rec_alpha}(c) and \ref{fig:rec_alpha}(d). 
The corresponding optimized interaction energies ($\epsilon_{\mu\nu}$) are plotted in Figs.~\ref{fig:rec_alpha}(e) and~\ref{fig:rec_alpha}(f). The values range approximately from $0$ to $3k_BT$. 
Given that typical contact probability numbers are very small, the optimized energies are just above thermal energy and are comparable to interaction energies of certain proteins. Exact values of the interaction parameters have been given in Table~S1 and S2 for the GM12878 and K562 cell lines, respectively.

In order to compare the normalization method introduced in the current work with the normalization procedure that is commonly used, namely the ICE technique, we have also carried out the IBD procedure on an ICE normalized matrix. More details of the ICE method that has been used here are given in the supporting material in section~S3. The ICE normalized contact matrix and the corresponding recovered matrix through IBD for both the cell line K562 and GM12878 are shown in Fig.~S3 of the supporting material. Clearly, the IBD method also recovers the contact probability matrix obtained with the ICE normalization. As will be discussed in further detail below, the normalization method has a significant effect on all the structural properties that have been evaluated in the current work.

\begin{figure}[t] 
	\begin{center}
		\begin{tabular}{c}
			{\includegraphics*[width=4.0in,height=!]{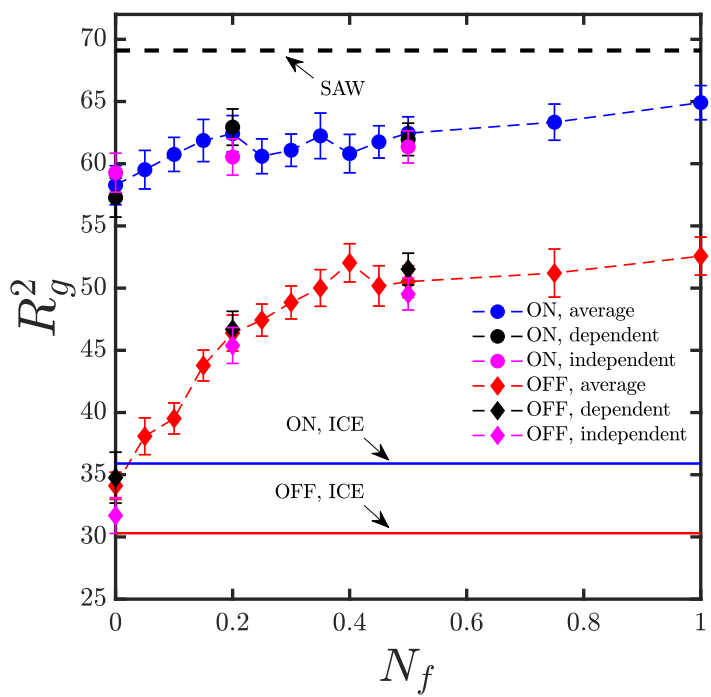}} \\
		\end{tabular}
	\end{center}
	\caption{Spatial extension of the polymer chain, quantified by the radius of gyration, $R_g^2$, computed at various values of the normalization parameter $N_f$ (see Eq.~\ref{eq:N_f} for definition), for both K$562$ (ON state) and GM$12878$ (OFF state) cell lines. All three coarse-graining techniques, i.e., dependent, independent and average, have been used. The black dashed line represents the value of $R_g^2$ for a chain executing self-avoiding walk statistics. Blue and red lines indicate the $R_g$ for ICE normalized ON and OFF state, respectively. 
	\label{fig:rg_nf} }
\end{figure}

The spatial extent of the chromatin polymer, as quantified by the square radius of gyration $R_g^2$, for different values of $N_f$ is presented in Fig.~\ref{fig:rg_nf}.  In the case of the cell line where the gene is ON (K562), the increase in $R_g^2$ for small values of $N_f$ is relatively less prominent and becomes nearly independent of $N_f$ as $N_f$ approaches one. It is clear that contact probabilities decrease with increasing $N_f$, since $N_s$ increases with $N_f$. It is consequently expected that with sufficiently large $N_f$, $R_g^2$ should approach the value for a self-avoiding walk. 
We have simulated a self-avoiding walk using the SDK potential with $\epsilon_{\mu\nu} = 0$; this represents a purely repulsive potential, and the result is shown as a black dashed line in Fig.~\ref{fig:rg_nf}.
In the cell line where the gene is OFF (GM12878), the value of $R_g^2$ increases relatively rapidly for small values of $N_f$ and reaches a nearly constant value for  $N_f \gtrsim 0.4$. However, the limiting value is significantly smaller than that of a self-avoiding walk. This suggests that some significant interactions are still present amongst the bead-pairs, even for $N_f$ approaching one. The influence of the different coarse-graining procedures was examined and it was found that the value of $R_g^2$ from all the three coarse-graining procedures agreed with each other within error bars (as seen from the data at $N_f=0, 0.2$ and $0.5$, for both the cell lines). This suggests that, at least as far as $R_g^2$ is concerned, the choice of coarse-graining method is not vitally important. 

However, the IBD results for ICE-normalized reference contact probability predicts a very different value for $R_g$ of the ON state (blue line) and OFF (red line) state. 
As can been seen, the $R_g^2$ for ON state using ICE normalization is close to the $R_g^2$ obtained here for OFF state at $N_f =0$. Interestingly this similarity is observed for many of the properties considered here, as will be discussed in more detail below.  

\subsection{\label{sec:3d_confi} Three-dimensional configuration of the $\alpha$-globin gene locus }

\begin{figure}[!hbt] 
	\begin{center}
		\begin{tabular}{c}
			{\includegraphics*[width=3.2in,height=!]{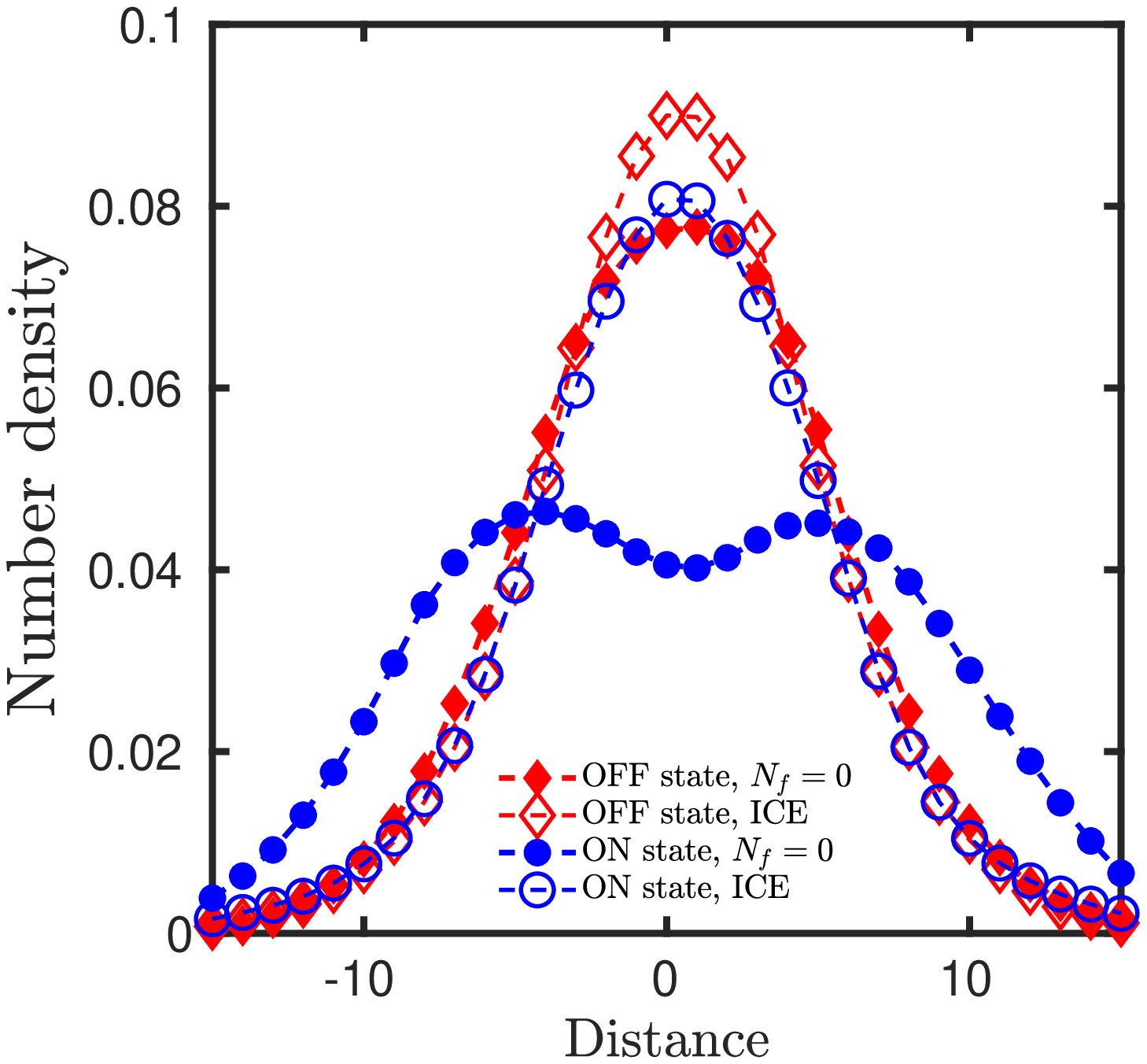}} \\
			(a) \\
		\end{tabular}
	    \begin{tabular}{c c}
	    	{\includegraphics*[width=3.2in,height=!]{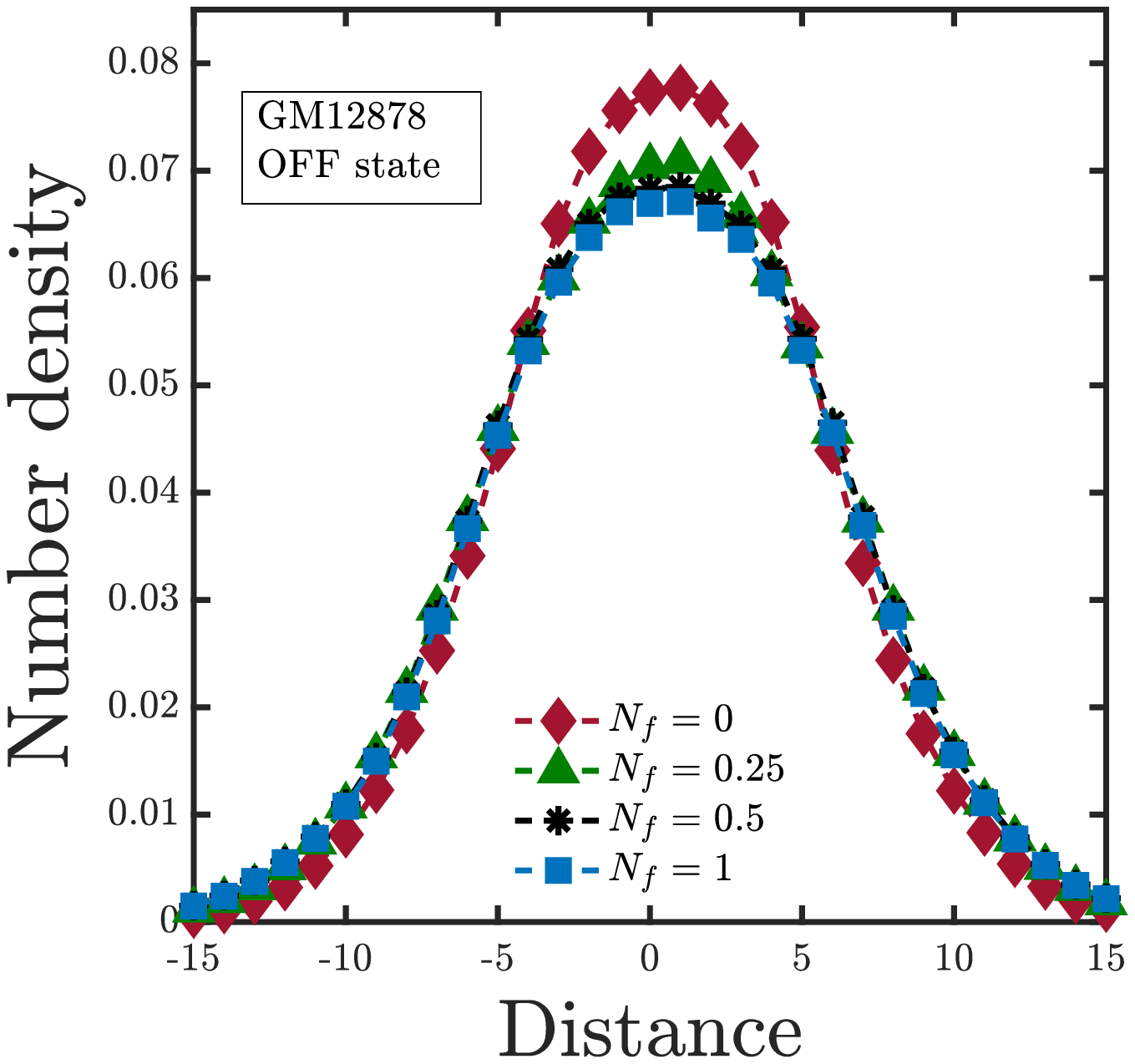}} &
	    	{\includegraphics*[width=3.2in,height=!]{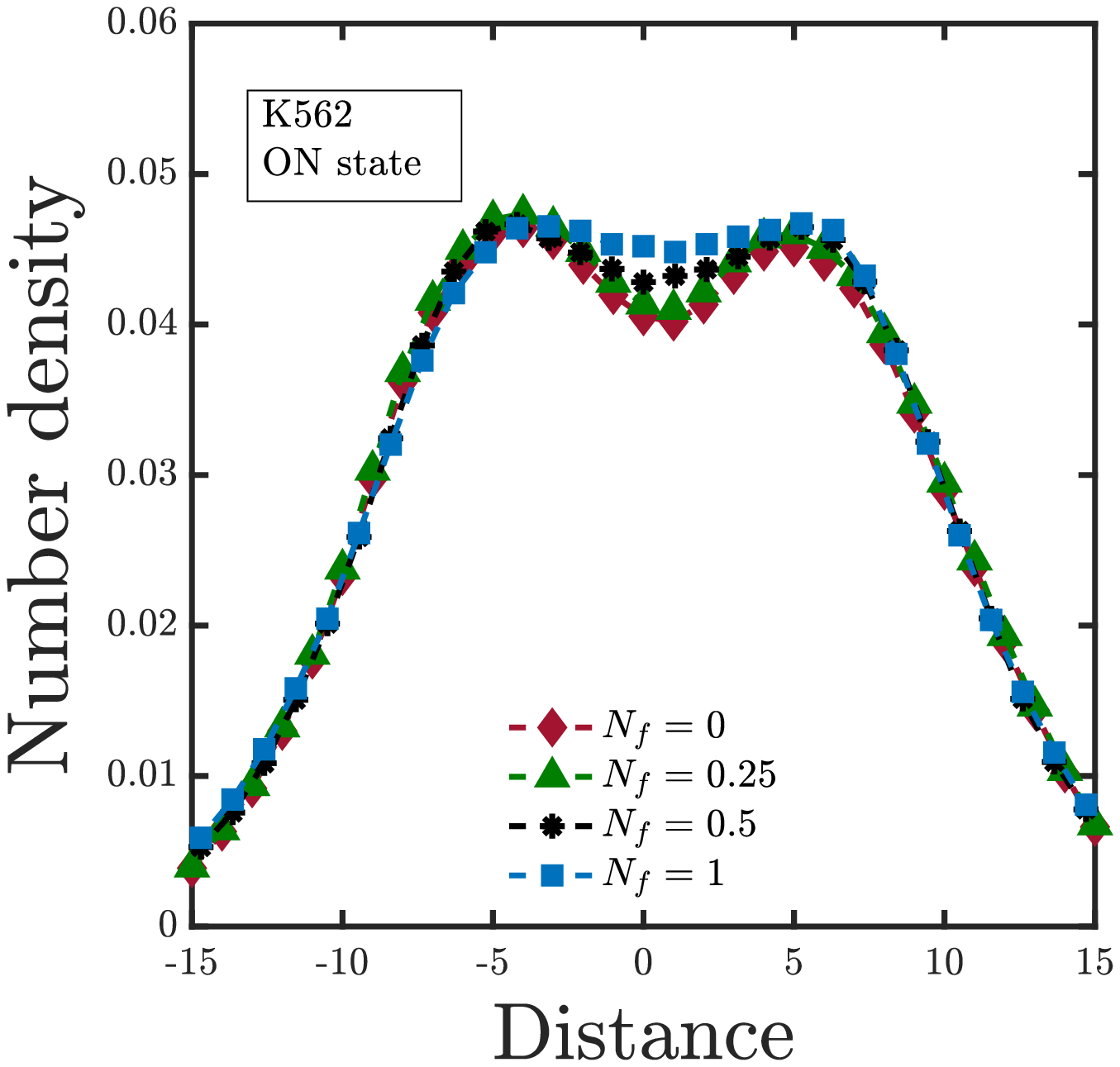}} \\
	    	(b)  & (c) \\
	    \end{tabular}
    \caption{
    	\label{fig:no_den} Comparison of the number density of beads along the major axis of the radius of gyration tensor, for various values of the normalization parameter $N_f$ (see Eq.~\ref{eq:N_f} for definition), (a) ON and OFF states at $N_f = 0$, (b) the OFF state, and (c) the ON state for various values of $N_f$.} 
	\end{center}
\end{figure}

\subsubsection{Shape functions}

Since chromatin folded in 3D can have spatial organization that is beyond simple spherically symmetric packing, various non-globular 3D shape properties (as described in section~\ref{sec:model}) have been analysed here. 

Eigenvalues of the radius of gyration tensor for polymer chains are usually reported in terms of ratios, either between individual eigenvalues, or with the mean square radius of gyration. For a chain with a spherically symmetric shape about the centre of mass, we expect $\langle \lambda^2_i \rangle /\langle R_\text{g}^2\rangle = 1/3$, for $i=1, 2, 3$, and $ \langle \lambda^2_i \rangle /  \langle \lambda^2_j \rangle = 1$ for all combinations  $i$ and $j$. For chain shapes with tetrahedral or greater symmetry, the asphericity $B = 0$, otherwise $B > 0$. For chain shapes with cylindrical symmetry, the acylindricity $C = 0$, otherwise $C > 0$. With regard to the degree of prolateness, its sign determines whether chain shapes are preponderantly oblate ($S \in \left[-0.25,0 \right]$) or prolate ($S, \in \left[0,2 \right]$). The relative anisotropy ($\kappa^{2}$), on the other hand, lies between 0 (for spheres) and 1 (for rods). 

All these properties are investigated for $N_f=0, 1$ and for the ICE normalization, and compared in the ON and OFF states, as displayed in Table~\ref{tab:shape}. It is clear that the while the chain is highly non-spherical in both states, it appears to be slightly more spherical in the OFF than in the ON state. 
The biggest difference is observed at $N_f=0$ between ON and OFF states. As we approach $N_f=1$, the difference between ON and OFF states is not so significant. With ICE, there is not much difference between the two states. As previously observed with the radius of gyration, ICE values are very close to the OFF state at $N_f=0$. 

\begin{table}[t]
	\setlength{\tabcolsep}{16pt}
	\renewcommand{\arraystretch}{1.1}
	\centering
	\caption{Various shape property based on the eigen values of gyration tensor $\bm{G}$ are defined here for $N_f$ = 0, $N_f = 1$ and ICE normalized contact matrix for K562 (ON state) and GM12878 (OFF state) cell line.}
	\label{tab:shape}
	\begin{tabular}{|c|ccc|ccc|}
		\hline
		\multirow{2}{*}{Shape properties} & \multicolumn{3}{c|}{K562 (ON state)}        & \multicolumn{3}{c|}{GM12878 (OFF state)}        \\ \cline{2-7} 
		& $N_f = 0$ & $N_f=1$ & ICE  & $N_f = 0$ & $N_f =1$ & ICE  \\ \hline
		$\langle\lambda_1^2\rangle/R_g^2$                     & 0.058     & 0.057 &  0.078       &  0.081  & 0.066  &  0.083        \\
		$\langle\lambda_2^2\rangle/R_g^2$                     & 0.164     & 0.175 &   0.189       & 0.201    & 0.177 &   0.195       \\
		$\langle\lambda_3^2\rangle/R_g^2$                     & 0.778     &   0.768 & 0.732         & 0.718     & 0.757 &  0.722        \\
		$\langle\lambda_2^2\rangle/\langle\lambda_1^2\rangle$ & 2.828     & 3.054  & 2.417          & 2.479    & 2.703 &   2.357       \\
		$\langle\lambda_3^2\rangle/\langle\lambda_1^2\rangle$ & 13.412     & 13.422  &   9.356       & 8.874   & 11.563 &  8.727        \\
		$B/R_g^2$                                             & 0.667     & 0.652  &   0.599       &  0.578   & 0.636 &    0.583      \\
		$C/R_g^2$                                             & 0.106     &  0.118  & 0.111         &  0.120   & 0.112 &   0.112       \\
		$S$                                                   & 0.913     & 0.816  &   0.988       &  0.772   & 0.926 &  0.867        \\
		$\kappa^2$                                            & 0.545     &  0.513  &     0.537      & 0.452     &0.525  &   0.497       \\ \hline
	\end{tabular}
\end{table}

\subsubsection{Density profiles}
\begin{sidewaysfigure}
	\centering
	{\includegraphics*[width=7in,height=!]{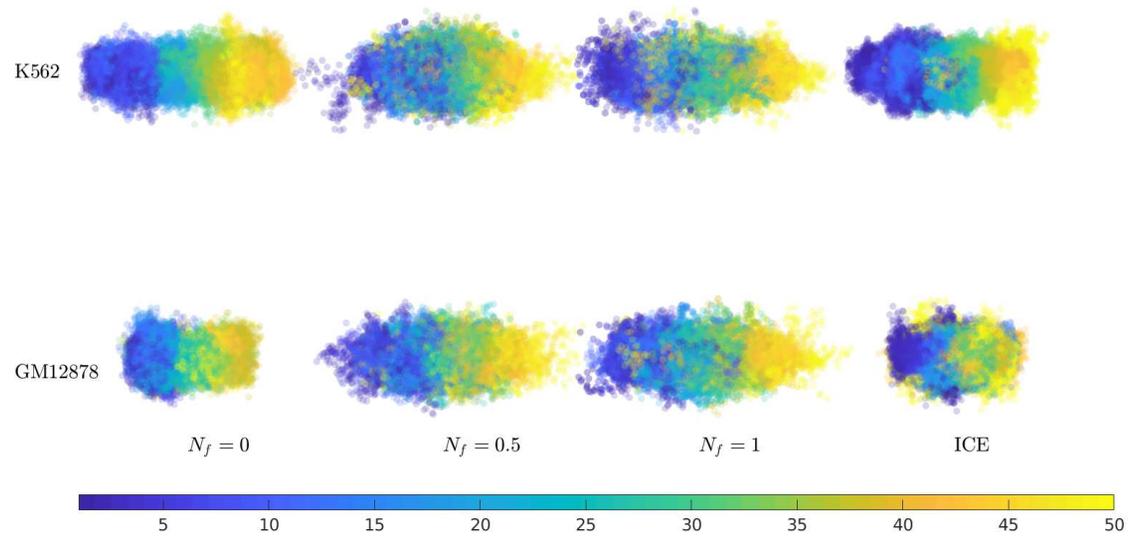}} \\
	\caption{
		\label{fig:3d_confi} Snapshots of 3D configurations, obtained by aligning chains along the major axis of the radius of gyration tensor and superimposing them on top of each other with transparency. Configurations at different values of the normalization parameter $N_f$ (see Eq.~\ref{eq:N_f} for definition) are displayed for cell lines K$562$ and GM$12878$. The colour assigned to each marker (blue to yellow) represents the bead number along the contour length (bead 1 to bead 50) of the polymer chain.}		
\end{sidewaysfigure}

To get a different prospective on the 3D organization of the gene, the density distribution about the centre of mass was considered. In order to do this, all polymer configurations were aligned along the major axis of the radius of gyration tensor $\textbf{G}$ and each bead position was binned and the number density of beads along the major axis was computed. As displayed in Fig.~\ref{fig:no_den}(a), in GM12878 (OFF state) cells, the number density shows a single peak at the center of mass position suggesting a symmetric organization around the centre of mass along the major axis. In the case of K562 (ON state) cells, the number density is seen to have a double peak, implying a bimodal distribution of polymer beads around the centre of mass along the major axis (Fig.~\ref{fig:no_den}(a)), as suggested by earlier 3D models for the $\alpha$-globin gene~\citep{Bau2011a,Paulsen2018}.  With an increase in $N_f$, a slight decrease in the number density at the core of the $\alpha$-globin gene in the OFF state is observed (Fig.~\ref{fig:no_den}(b)), while a decrease in extent of  bimodality is observed in the ON state (Fig.~\ref{fig:no_den}(c)).  
However, the differences for different $N_f$ values are less prominent at the peripherial regions of the globule.
Data comparing the density profiles for the three coarse-graining techniques (\textit{dependent}, \textit{independent} and \textit{average}) are provided in section~S4 of the supporting material. It was observed that the coarse-graining procedure did not have any influence on the density profiles.

We have also compared the density profile corresponding to the ICE-normalized matrix, displayed in Fig.~\ref{fig:no_den}(a) along with $N_f=0$. With the ICE normalization, both states (ON and OFF) show a single peak at the centre of mass. The bimodal nature of the ON state is no longer observed. This is a clear prediction that distinguishes the ICE-normalized result from the other results and can be tested in future experiments.

\subsubsection{3D conformations}
To obtain a snapshot of the 3D structure of the $\alpha$-globin gene locus, $1000$ different configurations from the ensemble were aligned along its major axis and then superimposed on top of each other, as displayed in Fig.~\ref{fig:3d_confi}, for both the cell lines at different  values of $N_f$ and with the ICE normalization. Each dot represents a bead and to make them visible, they have been made transparent to some degree. Different colors in the plot represent the bead number along the contour length of the polymer chain. As indicated from the shape functions and the density profiles, the snapshot shows that the structure is highly non-spherical in both cases, In particular, the K562 (ON state) cell line chromatin has a more extended configuration, with slightly higher density away from the centre of mass. 
As can be seen in Fig.~\ref{fig:3d_confi}, the snapshot for $N_f=0$ has some differences with snapshots for larger $N_f$ values. The value of $N_f$ was seen earlier to  affect average properties like $R_g^2$ (Fig.~4). The snapshots in Fig.~\ref{fig:3d_confi} show a similar behaviour as $R_g$ reflecting the variation for small $N_f$ and saturation for larger $N_f$. 

\subsection{\label{sec:c2d} 3D spatial distances and contact probabilities}

\begin{figure}[ptbh] 
	\begin{center}
		\begin{tabular}{c}
			{\includegraphics*[width=3.7in,height=!]{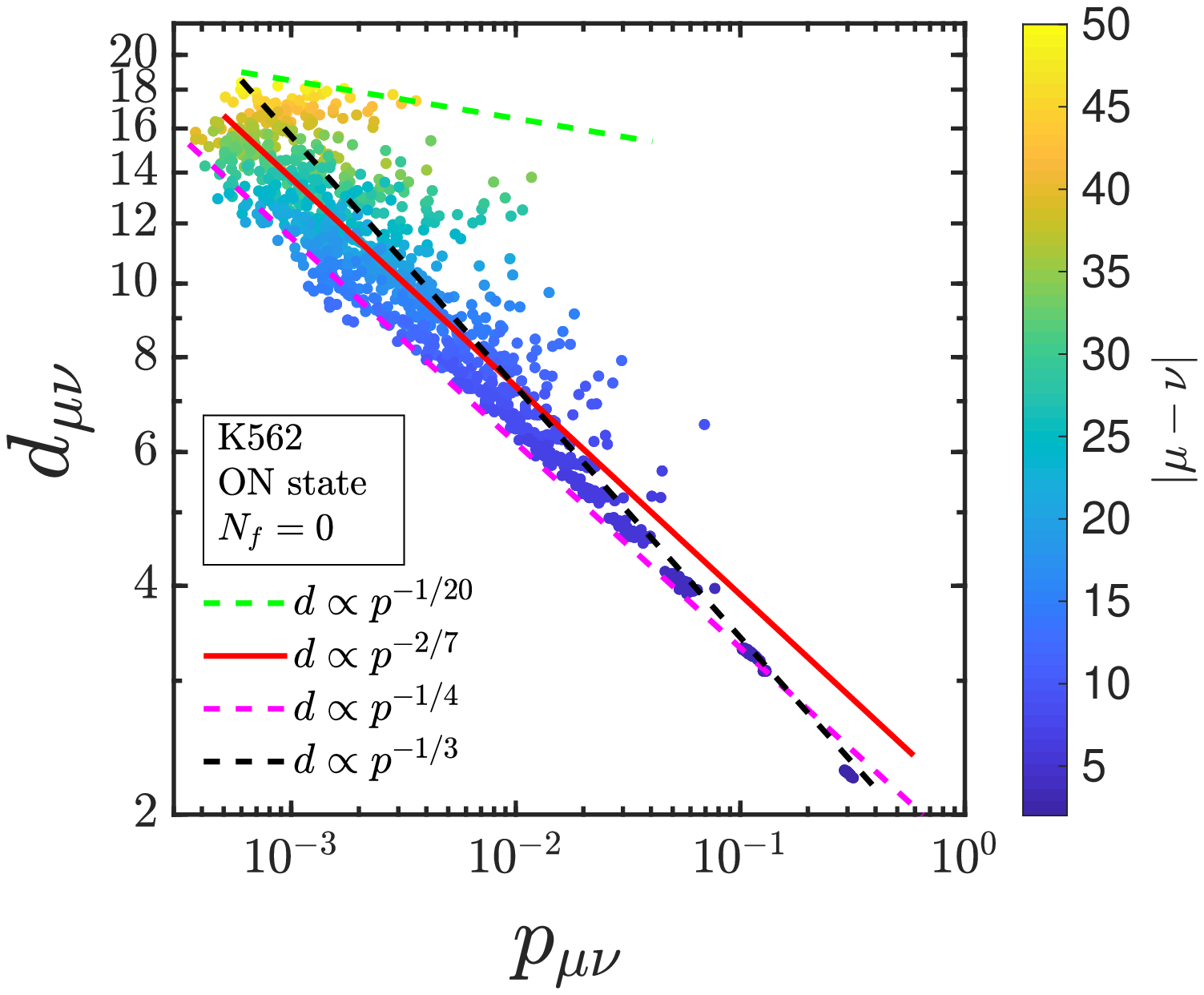}} \\
			(a) \\
			{\includegraphics*[width=3.7in,height=!]{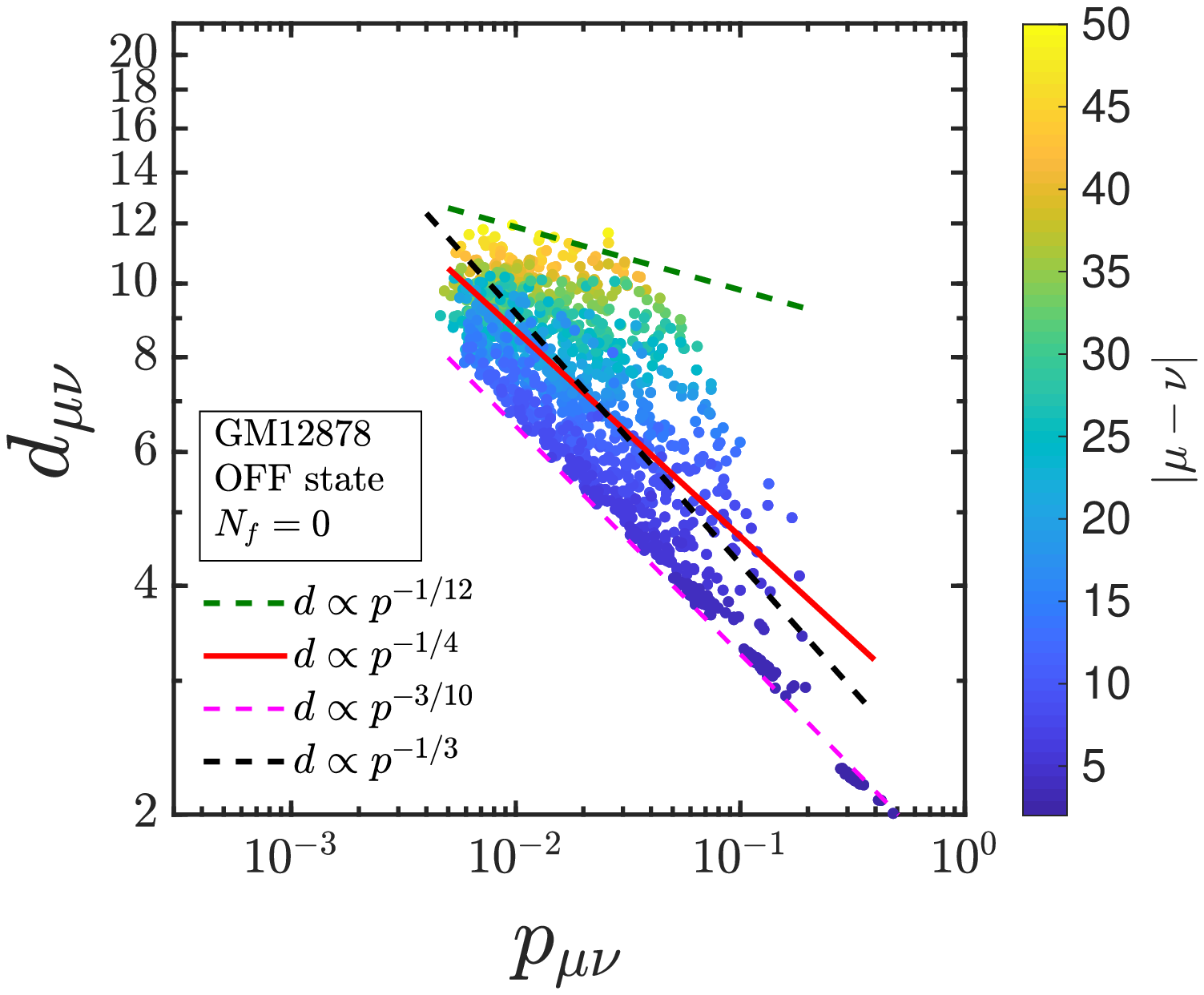}} \\
			(b) \\
		\end{tabular}
	\end{center}
	\caption{
		\label{fig:cp_dist} Dependence of mean 3D distances $d_{\mu\nu}$ on contact probabilities $p_{\mu\nu}$ for (a) K562 (ON state) and (b) GM12878 (OFF state) cell lines, respectively for $N_f=0$. For the K562 (ON state) cell line, the contact probabilities are bounded by power laws, $ d_{\mu\nu}\propto p_{\mu\nu}^\tau$, where $\tau$ varies from $-1/20$ (upper bound) to $-1/4$ (lower bound) as indicated by the green and magenta dashed line. Similarly, in the GM12878 (OFF state), $\tau$ varies from $-1/12$ to $-3/10$. The red line indicates the power law fitted to the simulation data points. The black dashed line represent the analytical relation between the contact probability and a spatial distance for an ideal polymer chain.}
\end{figure}

\begin{figure}[ptbh] 
	\begin{center}
		\begin{tabular}{c}
			{\includegraphics*[width=4.5in,height=!]{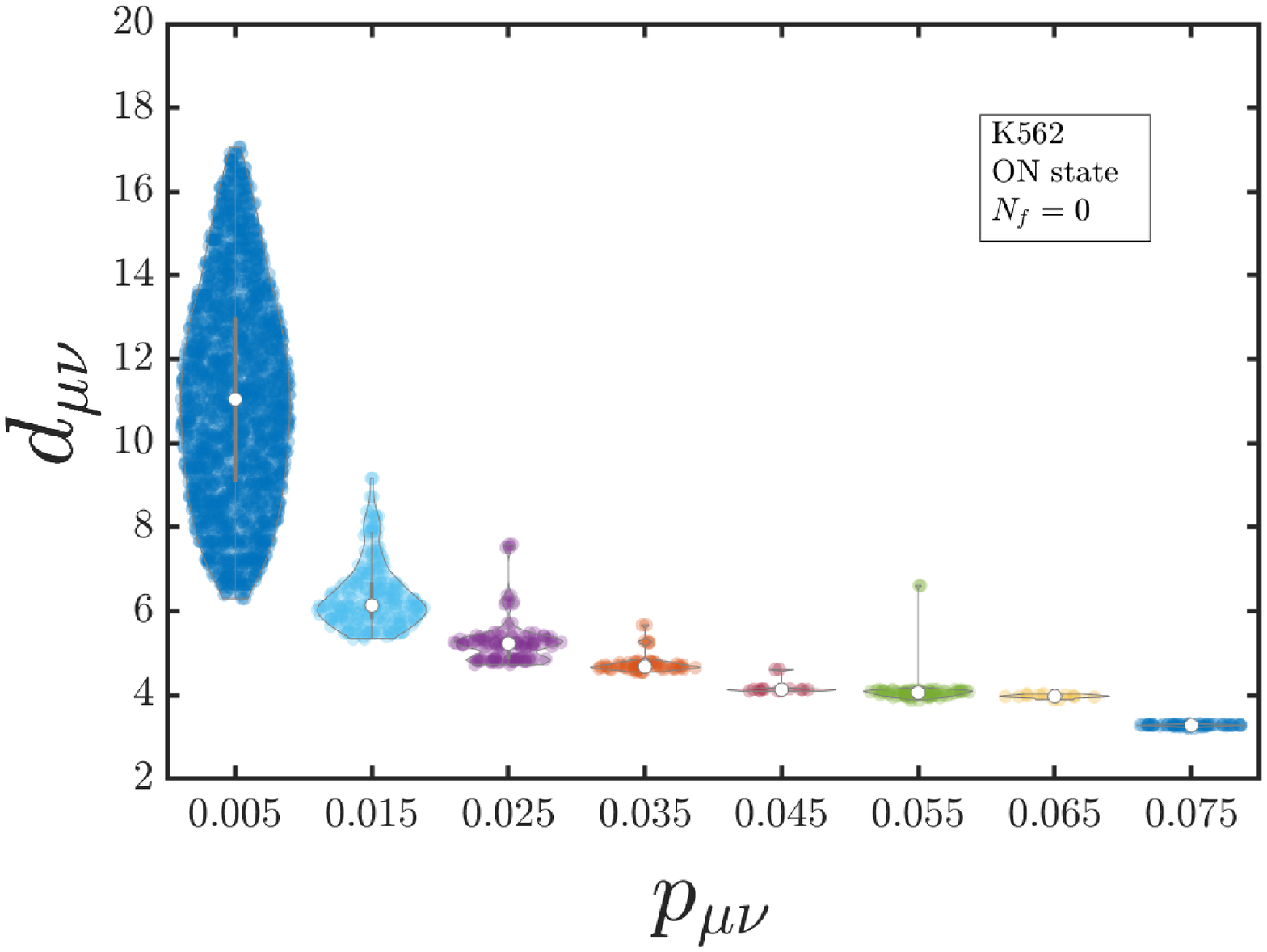}} \\
			(a) \\
			{\includegraphics*[width=4.5in,height=!]{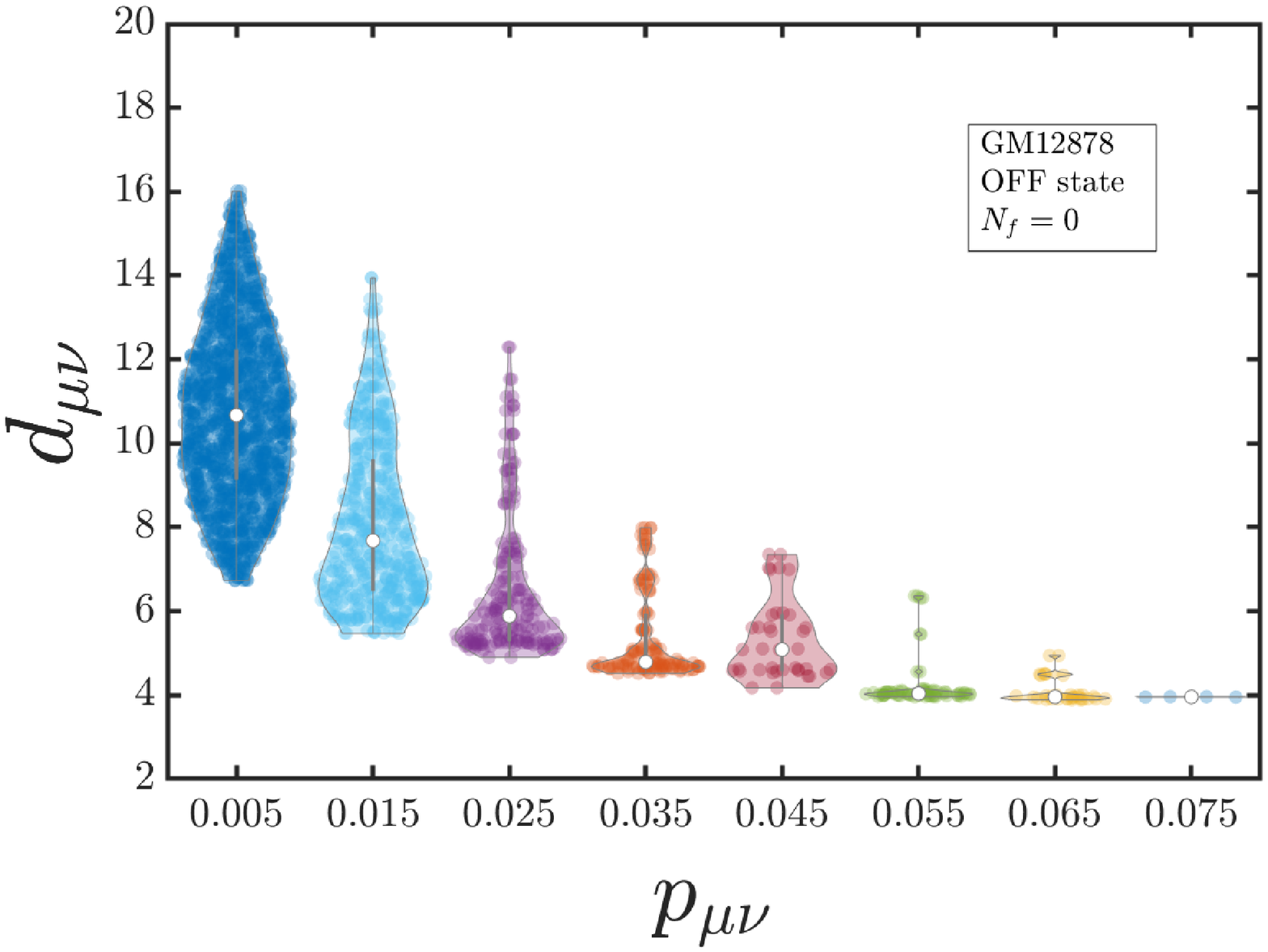}} \\
			(b) \\
		\end{tabular}
	\end{center}
	\caption{
		\label{fig:violin} Violin plots which display the probability distribution of mean 3D distances for selected ranges of contact probabilities in (a) the K562 (ON state) cell lines, and (b) the GM12878 (OFF state) cell lines for $N_f=0$.}
\end{figure}

The 3D conformation of the $\alpha$-globin gene locus has been investigated earlier \citep{Bau2011a,Paulsen2017}. These studies differ from the current work in some important aspects. Firstly, they assume that the contact counts between any two pairs can be converted to an equilibrium distance between those pairs through a certain pre-determined functional form. Secondly, instead of optimizing the interaction strengths to recover the contact counts, their simulations attempt to recover the equilibrium distances that have been derived from contact matrices. It is not clear in these cases whether the experimentally observed contact counts will be recovered by simulations. In this work, no assumptions have been made about the relationship between spatial distance and contact probability for any pair of beads. On the contrary in the present case, we can compute the spatial distances ($d_{\mu \nu}$) that are consistent with the contact probability matrix. Further, no configuration from the ensemble is discarded. 

The spatial distances calculated in the current work for the contact probabilities in the ON and OFF state are shown in Fig.~\ref{fig:cp_dist}(a) for K562 (ON state) and in Fig.~\ref{fig:cp_dist}(b) for GM12878 (OFF state) cell lines. Each point in these figures represents the ensemble-averaged 3D distance between a given pair of beads ($y$-axis) having a contact probability as indicated in the $x$-axis. As is immediately apparent, a wide range of 3D distances is possible, unlike what was assumed in earlier studies. It appears that the average 3D distance is not just a function of contact probability $p_{\mu\nu}$ (where the interaction between the beads plays a role), but is also a function of the distance along the contour between the beads ($\lvert \mu-\nu \rvert$) -- the color variation in Figs.~\ref{fig:cp_dist}(a) and (b) indicates the influence of contour length. The red line in both the figures are fitted power-laws to the data\. In both cases, the exponents are close to $-1/4$. But the interesting element here is the variability (scatter) in the data which shows that for a given contact probability value, there can be multiple values of 3D distances, with deviation of many units. 

To understand this variability better, we bin the same data and plot it as violin plots that display the mean 3D distance for a given small range of contact probabilities, as shown in Figs.~\ref{fig:violin}~(a) and~(b). It is clear that the distribution of points around the mean is very diverse -- bimodal in a few cases and with an extended tail in many cases -- suggesting that a simple functional form between the mean 3D distance and the contact probability may not be feasible. It must be reiterated here that many previous studies have assumed power law relations such as $d_{\mu\nu} \propto p_{\mu\nu}^\tau$ can be used, with exponents $\tau = -1$ \citep{Fraser2009,Duan2010}  and $\tau = -1/2$ \citep{rousseau2011three}, independent of $\lvert \mu-\nu \rvert$. Some groups have also assumed exponential  \citep{Tanizawa2010} and logarithmic decay of distance with probability \citep{Bau2011a}. As shown above, the results reported here do not support the usage of such simple functional forms. 
However, for an ideal chain, we know that contact probability $p \propto s^{-3/2}$ and the average 3D distance scales as $d\propto s^{1/2}$ where $s$ is the contour length between any two polymer beads. Combining these two, we get $d\propto p^{-1/3}$. This is shown by the black dashed line in Fig.~\ref{fig:cp_dist}. Clearly, the relation between mean 3D distance and the contact probability is significantly more complex than for a simple ideal chain. The relationship and its variability for $N_f=1$ and ICE normalization are discussed in the supporting material. In these instances as well,  the mean 3D distance is observed to be a function of both the contact probability and contour distance $\lvert \mu-\nu \rvert$.  

\section{\label{sec:conclusion} Conclusion}
The 3-dimensional organisation of chromatin based on publicly available chromatin conformation capture experimental data has been investigated. Unlike many existing models, the current work treats this as an inverse problem where interactions between different chromatin segments are computed such that the experimentally known contact probabilities are reproduced.
A polymer model and an Inverse Brownian dynamics (IBD) alogirthm has been developed for this problem which has the following advantages: (i) it does not assume any \textit{a priori} relation between spatial distance and contact probability, (ii) it optimizes the interaction strength between the monomers of the polymer chain in order to reproduce the target contact probability, and (iii) since hydrodynamics interactions are included, it is capable of investigating the dynamics of the chromatin polymer.

The main results of this work are as follows: (i) The IBD method was validated for a bead-spring chain comprising of 45 beads. It was observed that IBD reproduced the contact probability and the interaction strength (within 5\% of error), reflecting its reliability. (ii) Three different coarse-graining procedures -- \textit{independent}, \textit{dependent} and \textit{average} were used to map between the experimental and coarse-grained contact matrices. For the gene locus studied in this work ($\alpha$-globin gene), no significant differences between the three cases was observed, both for the gene extension and the density profile. (iii) A procedure for normalizing the contact count matrix was introduced with a parameter $N_f$ varying from $0$ to $1$, that reflected the two different extreme scenarios for estimating the sample size. For the GM12878 (OFF state), the gene extension increases rapidly initially with increasing $N_f$, while for the K562 (ON state) on the other hand (which is already in an extended state), there is a very little scope for further extension with increasing $N_f$. 
(iv) We also simulated the K562 and GM12878 cell line data with ICE-normalization. Structural properties such as shape properties, density profile and 3D configuration show an significant difference between ICE and other normalization technique.  
(v) Since there is a relationship between the normalization method (value of $N_f$ or ICE) and physically measured properties such as the radius of gyration, it is conceivable that the appropriate normalization method can be inferred from experiments such as FISH, Chip-seq, etc. 
(vi) The structural properties of the $\alpha$-globin gene locus were investigated in terms of shape functions, bead number density distributions, and 3D snapshots. In the ON state (K562), $\alpha$-globin appears to lack any prominent interactions, and exists in an extended structure. Whereas in the case of GM12878 (OFF state), the gene appears to be in a folded state. This is also consistent with theory, as in the ON state (K562) the transcription factors need to access the gene, while the structural status of the OFF state (GM12878) should be to avert the transcription factor, resulting in gene silencing. 
(vii) The density profile along the major axis of the radius of gyration tensor also supports the extended structure in cell line K562 (ON state) and a sharp cluster of monomers at the core of GM12878 (OFF state). 
(viii) The dependency of spatial distance on contact probability has been investigated, and it is shown that the usage of simple functional relationships may not be realistic. 
(ix) No bimodal nature was observed in the density profile of ON state with ICE normalization. Both ON and OFF state shows a single peak at the center of mass indicating a collapsed globule.

Most of the results in this work are  predictions that may be tested in suitably designed experiments. We predict that the spatial segmental distance is not only dependent on the contact probability but also on the segment length along the contour. One of the ways to test our prediction is to perform 3D FISH on segment-pairs having the same contact probability but different segment length. A difference in distance obtained from the FISH experiment will validate the predictions made in the current work. Shape properties and density profiles of the $\alpha$-globin locus are also predicted and can be tested using techniques like super-resolution microscopy and electron microscopy.
We require these additional experiments to determine the appropriate normalization.
Our work predicts that 3D distances, shape properties, density profile etc. will depend on the precise nature of normalization. Hence, the appropriate normalization methodology may be determined by comparing our results with future experiments that measure these quantities.

One of the concerns regarding our work could be that this study simulates only a short segment. However most of the biologically relevant processes happen in the length scale of a gene (or a few genes). Hence, it is essential to zoom-in and study the organisation and dynamics of short segments. Given that chromatin is organized into small local domains (TAD/chromatin domains) having only local interactions predominantly, it may be reasonable to analyse one locus/domain at a time. The IBD algorithm can also be used to study the static and dynamics properties of the whole genome by considering a longer polymer chain. Several sampling techniques can be utilized to sample the phase-space efficiently such as parallel tempering techniques~\cite{Bunker2000ParallelEV}. This method can be used to check the validity of the simplest model for a given contact probability matrix. In other words, if a model does not converge to the desired probabilities even after proper sampling, it implies that the model (as represented by the Hamiltonian or the included physics) may require modification, and a more sophisticated model may be required. For instance, we have chosen the simplest model that can reproduce the experimentally observed contact probability map. A lack of convergence (even after proper sampling) may imply the need for adding additional physics into the model. For example, certain far away contacts may require the addition of non-equilibrium processes like loop extrusion. Since we use Brownian dynamics, our model can be extended to incorporate such non-equilibrium processes. 

Since the model has dynamics, with hydrodynamics interactions built in, it has the potential to be used to address problems in the future involving dynamics of the 3D chromatin polymer between different chromatin states. Currently in this model only chromatin configuration capture data has been considered. However, the model may be extended to incorporate more data (histone modification data, CHIP-Seq data of certain proteins) and address chromatin organization on the length scale of genes in more detail. 
Recent experiments suggest that 3D chromatin organization is driven by two different dynamic processes namely, phase separation and loop extrusion. Since our model is capable of studying dynamics, the model may be extended to investigate the interplay between different dynamic processes in determining chromatin organization. 
With the capability of analysing the 3D configuration along with chromatin dynamics, IBD can complement experimental research and also provide deeper and more useful insights based on the same.

\section*{Author contributions}  
K.K., B.D., R.P., and J.R.P. designed the research. K.K. wrote code, carried out simulations and analysed the data. K.K., B.D., R.P., and J.R.P. wrote the manuscript.

\section*{Acknowledgements} We appreciate the funding and support from the IITB-Monash Research Academy, and from SERB, DST India via grant number  EMR/2016/005965. We gratefully acknowledge the computational resources resources provided at the NCI National Facility systems at the Australian National University through the National Computational Merit Allocation Scheme supported by the Australian Government, the MonARCH facility maintained by Monash University, and by IIT Bombay.

\bibliography{biophysics}
\end{document}